\documentclass[aps,pre]{revtex4}

\usepackage{graphicx}
\usepackage[utf8]{inputenc}
\usepackage[english]{babel}
\usepackage{amsmath}
\usepackage{enumerate}
\usepackage{epstopdf}
\usepackage{amssymb}

\begin{document}

\title{The Role of Data in Model Building and Prediction: A~Survey Through Examples}

\author{Marco Baldovin $^{1}$, Fabio Cecconi $^{2}$, Massimo Cencini $^{2}$, Andrea Puglisi $^{3}$ and  Angelo Vulpiani $^{1,4,}$*}  

\affiliation{%
$^{1}$ Dipartimento di Fisica, ``Sapienza'' Universit\`a di Roma, p.le A. Moro 2, 00185 Roma, Italy; baldovin.m@gmail.com\\
$^{2}$ Istituto dei Sistemi Complessi, CNR, via dei Taurini 19, 00185 Rome, Italy; fabio.cecconi@roma1.infn.it (F.C.); massimo.cencini@cnr.it (M.C.)\\
$^{3}$ CNR-ISC and Dipartimento di Fisica, Sapienza Universit\`a di Roma, p.le A. Moro 2, 00185 Roma, Italy; andrea.puglisi@roma1.infn.it\\
$^{4}$ Centro Linceo Interdisciplinare ``B. Segre'', Accademia dei Lincei, via della Lungara 10, 00165 Rome, Italy} 

\date{\today}

%

\begin{abstract}
 The goal of Science is to understand phenomena and systems
  in order to predict their development and gain control over them.  In
  the scientific process of knowledge elaboration, a crucial role is
  played by models which, in the language of quantitative sciences,
  mean abstract mathematical or algorithmical representations.
  This short review discusses a few key examples from
    Physics, taken from dynamical systems theory, biophysics, and
    statistical mechanics, representing three paradigmatic procedures
    to build models and predictions from available data. In the case
    of dynamical systems we show how predictions can be obtained in a
    virtually model-free framework using the methods of analogues, and
    we briefly discuss other approaches based on machine learning
    methods.  In cases where the complexity of systems is challenging,
    like in biophysics, we stress the necessity to include part of the
    empirical knowledge in the models to gain the minimal amount of
    realism. Finally, we consider many body systems where many
    (temporal or spatial) scales are at play---and show how to derive
    from data a dimensional reduction in terms of a Langevin dynamics
    for their slow components.
\end{abstract}

\maketitle

\section{Introduction}

It is not an exaggeration to say that models are unavoidable in
scientific practice, and that it is impossible to have real science
without them. Models are representations and/or abstractions of real
phenomena \cite{Wiener,morgan1999models} that allow for predictions on
the phenomena and for understanding the impact of changing some
component or parameter of the system. They can be material, such as
scaled models of an airplane in a wind tunnel adopted in engineering
or simple organisms used in biology (e.g.,~Drosophila or bacteria in
evolutional studies). Other models, the subject of the present review,
can be expressed in mathematical terms such as equations or
algorithms.  Even those very complete and elegant descriptions that we
call \emph{theories}, such as classical mechanics or electrodynamics,
are nothing but very sophisticated models.

A good starting point for a general discussion about the role of
mathematical models for the description of phenomena is {\it A
  dialogue on the application of mathematics} written by A. R\'enyi
\cite{ren}.  The~protagonists are King Hieron and Archimedes, whose
burning mirrors allowed the Syracusans to sink half of the Roman fleet
effortlessly.  In the dialogue Archimedes illustrates the applications
of mathematics to concrete problems, the central role of models, and
the thoughtfulness necessary to master the art of constructing them.

\begin{quote}
First of all, one can construct many mathematical models for the same
practical situation, and one has to choose the most appropriate, that
which fits the situation as closely as practical aims require (it can
never fit completely). At the same time, it must not be too
complicated, but still must be mathematically feasible. These are, of
course, conflicting requirements and a delicate balancing of the two
is usually necessary.  [\dots] You have to approximate closely the real
situation in every respect important for your purposes, but lay aside
everything which is of no importance for your actual aims. A model
needs not to be similar to the modeled reality in every respect, only
in those which really count. On the other hand, the same mathematical
model can be used to fit quite different practical situations.
[\dots] In trying to describe such a complicated situation, even a very
rough model may be useful because it gives at least qualitatively
correct results, and these may be of even greater practical importance
than quantitative results. My experience has taught me that even a
crude mathematical model can help us to understand a practical
situation better, because in trying to set up a mathematical model we
are forced to think over all logical possibilities, to define all
notions unambiguously, and to distinguish between important and
secondary factors. Even if a mathematical model leads to results which
are not in accordance with the facts, it may be useful because the
failure of one model can help us find a better one \cite{ren}.
\end{quote}

The point of view of Archimedes was shared by  Rosenblueth and Wiener \cite{Wiener}
who concluded their paper ``on the role of models in science'' writing:

\begin{quote}
The ideal formal model would be one which would cover the entire
universe, which would agree with it in complexity, and which would
have a one to one correspondence with it. Any one capable of
elaborating and comprehending such a model in its entirety, would find
the model unnecessary, because he could then grasp the universe
directly as a whole. He would possess the third category of knowledge
described by Spinoza. This ideal theoretical model cannot probably be
achieved.  Partial models, imperfect as they may be, are the only
means for understanding the universe. This statement does not imply an
attitude of defeatism but the recognition that the main tool of
science is the human mind and that the human mind is~finite.
\end{quote}

It is pretty impossible to detail the different procedures which had
been followed to build the many models used in science, also because
it can be safely claimed that there are not systematic protocols for
model building.  Possibly a lucky non-trivial exception is given by
the framework of classical physics where a fairly clear approach
exists. Once the forces are understood, one can write down proper
differential equations that may be difficult to solve, but that always
allow us to obtain reliable and useful results, for instance by means
of qualitative or numerical analysis. Another example is when a
reduced model is derived from a more general theory, for instance this
the case of the Lorenz's model \cite{lor63}, which is obtained from a
rather crude simplification of fluid dynamics equations. Such a model,
in spite of its (apparent) simplicity, played a key role by allowing
us to realize that many irregular motions encountered in nature can be
due to chaos. This is a clear example of how even a ``partial model''
can help for understanding natural phenomena.

The above approach is decidedly not applicable to model biological
phenomena, medical or economical issues, not to mention social
behaviors. In these areas there is nothing like Newton's or Maxwell's
laws, and therefore a non trivial model can only be
  achieved either by a pragmatic approach using proper variables that
  can be guessed only via a good understanding of the subject~\cite{tacchella2018} or from some (often profound) intuition,
usually suggested by empirical observations and/or analogies. An~instance of the latter is the model of Lotka-Volterra
\cite{volterra1927,lot} which was built in analogy with kinetic theory
to describe the evolution of simple ecosystems.

Given the rising pace at which data are produced and accumulated, we
think it is timely to briefly review the different approaches to
employ available data to build models and predictions of certain
phenomena, discussing both classical and modern approaches. To
illustrate the possible strategies, we will discuss some specific
examples, taken from dynamical systems, the physics of biopolymers and
statistical mechanics of many particle systems.

The paper is organized as follows.  Section \ref{sec:interlude} is
devoted to a short interlude about the Lorenz and Lotka-Volterra
equations, which had an important role in the modeling of natural
phenomena and are here presented mainly for historical reasons. Then
the discussion will focus on the role of data in the process of
building models and formulating predictions.  Section
\ref{sec:analogues} is devoted to the use of data and models for the
prediction problem.  In Section \ref{sec:protein} we describe the data
driven approach for protein modeling.  Section \ref{sec:langevin}
treats the procedures to build Langevin equations, also from data, in
the case of systems with fast and slow variables.  In Section
\ref{sec:conclusion} some general remarks about the relevance of the
used variables, the proper coarse graining level, and open problems in
the modeling of phenomena using just data.

\section{An Interlude: The Story of Two Seminal Models\label{sec:interlude}}
It is instructive to start with a short interlude on the genesis of
two models which had a very important role in the history of modeling,
and show the role of intuition, as well as mathematics, in the building
of a description of natural phenomena.  The aim of such models
is not a detailed representation of the problem under investigation, but
to catch some (qualitative) aspects.

\subsection{The Lorenz's Model: How to Obtain Something Interesting 
with a Crude Assumption\label{sec:lorenz}}
In his famous 1963 paper \cite{lor63}, Lorenz, studying the problem of
atmospheric convection, derived a low dimensional model which is one
of the first examples of deterministic system with chaotic behavior.
Let us briefly recall the used approach.  He had to face a rather
common problem: Given a nonlinear partial differential equation
\begin{equation}
\partial_t \psi({\bf x},t) = {\cal L}[\psi({\bf x},t), \nabla\psi({\bf x},t), \Delta\psi({\bf x},t)]   \label{B1}
\end{equation}
where $\psi$ is a vector field, we want to find a set of differential
equations that approximate (\ref{B1}).  In the case treated by Lorenz
$\psi = ({\bf u}, T)$, where ${\bf u}$ and $T$ denote the velocity and
temperature fields respectively, and the evolution law (\ref{B1}) is
the Boussinesq equation which can be considered the ``true''
description (i.e., the theory) of the system under consideration,
i.e., atmospheric convection.

In general, Equation~(\ref{B1}) cannot be solved analytically,
thus one is forced to resort to a numerical approach. To do this, the
first step is to transform Equation~(\ref{B1}) into a set of ordinary differential
equations. A customary procedure (the so-called Galerkin method)
consists in approximating $\psi({\bf x},t)$ in the form
\begin{equation}
\psi({\bf x},t)=\sum_{n<N} a_n(t) \phi_n({\bf x}),
\label{B2}
\end{equation}
where $\{\phi_n\}$ are suitable orthonormal, complete
functions. Substituting (\ref{B2}) in (\ref{B1}), one obtains a set of
differential equations for the coefficients $\{a_n\} $:
\begin{equation}
\frac{d a_n}{dt}=F_n(a_1,a_2,.., a_N)\,\,\, , \,\, n=1,2,...,N.
\label{B3}
\end{equation}

For the above system to be a good approximation of (\ref{B1}), $N$ has
to be very large.  For instance, in applications in meteorology or
engineering, the value of $N$ easily reaches $10^9$ and even more. It
is worth mentioning that in the Galerkin method, from a mathematical
point of view, the choice of the functions $\{ \phi_n \}$ is largely
arbitrary.  The unique criterion is that in the limit $N\gg1$ the
truncated Equation (\ref{B3}) is a good approximation of the original
Equation (\ref{B1}).  In order to go beyond simple choices, such as
trigonometric functions, if one wants to avoid the use of very large
$N$ (this in principle is possible if the dimension of the system is
not huge) it is necessary to use a ``clever'' complete, orthonormal set
of eigenfunctions, which can be obtained only using some experimental
data. An example of such procedure---the proper orthonormal
decomposition (POD) \cite{Holmes2012}---is sketched in the Appendix.

In his celebrated paper, Lorenz  used  $N = 3$: In
particular, $2$ harmonics for the temperature and $1$ for the velocity. 
Its aim was to explore whether such a drastic approximation could reveal
some qualitative features of the original problem. His
famous, apparently simple, model consists of the following three ordinary differential equations:
\begin{equation}
\frac{dx}{dt}= -\sigma x + \sigma y, \,\,\,
\frac{dy}{dt}= -xz+rx-y, \,\,\, 
\frac{dz}{dt}= xy -bz,
\label{B4}
\end{equation}
where $(x, y, z)$ are proportional to $(a_1, a_2, a_3)$, and $\sigma$,
$b$ and $r$ are constants related to the properties of the fluid; in
particular, $r$ is proportional to the Rayleigh number.

It turns out that even the simplified, and physically very crude, Equation (\ref{B4})
cannot be solved analytically. The numerical study performed by Lorenz
showed a highly non-trivial behavior, characterized by non-periodic
evolutions and, more remarkably, by the property that even very
close initial conditions quickly (exponentially fast) evolve in time into
very different states. This is the essence of deterministic chaos.

So the importance of Lorenz's model lies in having shown that it is
possible to obtain a chaotic (irregular) behavior even in
low-dimension deterministic systems.  This result finally led to
understand that the complexity of the temporal evolution occurring in
turbulent fluids is not a mere superposition of many elementary events
(say, many Fourier harmonics), but originates from the nonlinear
structure of the equations, which gives rise to chaos
\cite{Ruelle1971}.

\subsection{The Lotka-Volterra Model: The Power of the Analogy\label{sec:volterra}}
Volterra came into contact with data from fishing for the period 1903--1923 gathered by his future
son-in-law Umberto D'Ancona. In particular, the data concerned the
presence of cartilaginous fish in the catch of three Adriatic ports:
Trieste, Venice, and Fiume (now Rijeka).

 Volterra simplified in a drastic way the description of marine
 ecosystem considering only two variables: $x$ and $y$ representing
 the number of prey (small fishes) and predators (big fishes),
 respectively.  Then he reasoned by analogy with the kinetic theory:
 \emph{The big fish ``collide'' with the small fish and with a certain
   probability the former eats the latter}, and thus came to identify
 two differential equations (see below) as a model capable of
 accounting for D'Ancona's observations \cite{volterra1927}.

Some years before, Lotka \cite{lot} observed experimentally some
patterns in certain chemical reactions, but these patterns turned out
to be oscillatory \cite{bac}.  Volterra, unaware of the Lotka's work,
arrived to the same equations for the dynamics of the two populations:
\begin {equation}
\frac{dx}{dt} = ax - bxy \, \,, \, \,
\frac{dy}{dt} = -cy+ dxy, \, \, \label{eq:LV}
\end{equation}
where   the constants $a$, $b$, $c$, and $d$ are positive. The
linear terms are quite transparent: Assuming unlimited food
resources, the absence of predators leads to an exponential
proliferation of the prey; analogously, in the absence of prey,
predators become extinct. In the fish-related case that aroused
Volterra's interest, the non-linear terms have  the following
interpretation: When both cartilaginous fish and their prey increase,
the population transfer from prey to predators also increases.

Despite the simplicity of the model,  it is
possible to derive a theoretical prediction that is anything but
trivial, and in particular the \emph{mathematical explanation} of
zoological observations. Suppose the two fish populations are not in
equilibrium. Then the abundance of small fish leads to an increase in
the population of cartilaginous fish. These, in turn, by hunting will quickly
cause a decrease in the population of the other species, until some
cartilaginous fish will starve, allowing the repopulation of small
fish. And so on, periodically.

Perhaps some readers will have noticed in Equation \eqref{eq:LV} the
(bilinear) structure similar to that of Boltzmann equation for kinetic theory of gases. This is
not a mere coincidence: The nonlinear terms were introduced by Volterra
noting the analogy between the prey/predator interaction and the
impact of two atoms in the kinetic theory while, as we have seen
above, Lotka had chemical reactions in mind---in fact, it is the same
mechanism.

Once the model based on the Equation \eqref{eq:LV} has been
constructed, we can reason about it in a purely mathematical terms,
asking ourselves if it is possible to extend the predictive ability of
the model itself. For instance, it is natural to wonder if the model
only ``works'' with algebraic nonlinearities, or if we can consider
the case with $N$ different species $x_1$, \dots, $x_N$. In ecology there is nothing similar to Newton's
mechanics, and therefore such generalizations cannot be sought by
relying on first principles. However, the analogy with the
Lotka-Volterra model and the consistency with the relevant ecological
facts emerge as very natural constraints. Thus, Kolmogorov introduced
a generalization of the Lotka-Volterra equations for the $N = 2$,
while Smale considered the case with $N \ge 3$ for a class of possible
equations \cite{murray2001}.

Limiting ourselves to quadratic nonlinearities, the obvious generalization of the Lotka-Volterra model is
\begin{equation}
\frac{dx_n}{dt} = a_n x_n (1 - \sum_{j = 1}^N b_{n, j} x_j), \label{eq:4}
\end{equation}
where $a_n$ is positive for prey (herbivores) and negative for
predators; the diagonal terms describe the competition between
individuals of the same species, the non-diagonal ones specify the
type of interactions between the various species, for instance of
parasitic, mutualistic, or prey/predator type.

Summarizing: Data, or rather observations, were fundamental for both
Lotka and Volterra, but the brief accounts of their motivations and
their arguments for choosing Equation \eqref{eq:LV} suggest a
predominantly heuristic role. Analogy, mathematical intuition and
deduction are the main components in the justification of the
equations of the model, which both make an effort to show to be
\emph{consistent} with fundamental ecological facts. This provides the
theoretical basis to derive, this time from the model and therefore
mathematically, new conjectures to be submitted to experimental
observation. We conclude remarking that generalization of the
Lotka-Volterra equations such as (\ref{eq:4}) are still widely used to
study the evolution of ecosystems and in population biology \cite{murray2001}.

\section{The Role of Data and Models for the Prediction\label{sec:analogues}}
Forecasting the future has always been a desire, and a need,
for humans and a natural motivation for science and the development of
models. Assuming that the system of interest can be accessed
experimentally, having a time record of its evolution, how can we
predict its future evolution? The~problem so stated is too general and
vague. In order to make some precise statement we have to define the
classes of system of interest.

Here, we shall restrict our interest to deterministic dynamical
systems, whose future is uniquely determined by their present state
$\bm x=(x_1,\ldots,x_N)$, where $\{x_i\}_{i=1}^{N}$ are all the
variables necessary to describe the system state, whose evolution is ruled
either by discrete time maps,
\begin{equation}
  \bm x(t+1)=\bm F(\bm x(t)) \,,
\label{eq:map}
\end{equation}
or by ordinary differential equations,
\begin{equation}
  \frac{d\bm x(t)}{dt}=\bm f(\bm x(t)) \,.
\label{eq:ode}
\end{equation}

In this framework, we can distinguish two cases: When we know and can
measure the whole state vector $\bm x$ and thus have access to a
sequence of its states; when we only have a time record
of some scalar quantity related to the system variables.

\subsection{Prediction When We Know the System State Vector\label{sec:pred1}}
This case is actually rather infrequent.  However, assuming that we
are in this fortunate circumstance we have two possibilities.  The
first corresponds to the ideal scenario in which we know the laws
ruling the evolution of the state variables, i.e., we know the right
hand side of Equations~(\ref{eq:map})~or~(\ref{eq:ode}). The second case is
when we do not have such a knowledge.

When the evolution laws are known  the main limitation to the possibility
of predicting the future is due to the inaccuracy in our knowledge of the
present state---\textit{the initial condition}---of the
system. Indeed, it is by now well established that the ubiquity of
chaos in nonlinear dynamical systems leads to a sensitive dependence
on the initial condition.  In mathematical terms, this means that even
if we precisely know the r.h.s. of Equations~(\ref{eq:map}) or
(\ref{eq:ode}), if the equations give rise to a chaotic dynamics, any
small uncertainty, $||\delta \bm x(0)|| \ll 1$, on the system state at time $t=0$
will grow exponentially in time, $||\delta \bm x(t)|| \approx ||\delta
\bm x(0)|| \exp{(\lambda t)}$, at a rate given by the maximal Lyapunov
exponent, $\lambda>0$, of the system.  In this case, knowing the past
evolution of the system does not add much to the prediction, the best
we can do is simply to have the most accurate measurement of the state
at a given time, i.e., $||\delta \bm x(0)||=\delta_0 \ll 1$, and then solve
the equations to predict the behavior knowing that if we accept an
accuracy $||\delta \bm x(t)|| \approx \Delta$ on the state, the
prediction will be valid up to a time of order \cite{libro}
\begin{equation}
T(\Delta,\delta_0)\approx \frac{1}{\lambda}\ln\left(\frac{\Delta}{\delta_0}\right)\,,\label{eq:predtime}
\end{equation}
where the above equation holds under the assumption that not only
$\delta_0$ but also $\Delta$ is small, so that the uncertainty is
effectively controlled by the linearized dynamics.

More interesting is the case in which we do not know the equations of
motion: In this case we can either hope to build the model, i.e., derive
the equations using the time record of the system state and then use
it to make predictions (facing the difficulties of the previous case),
or attempt a model-free~prediction.

Clearly deriving the equations of motion from data records without
inputs from theories or intuition, i.e., virtually in an automated way,
would be the grand goal of data driven science, i.e., a sort of
reverse-engineering from data to infer predictive mathematical
models. This would be particularly useful in biophysics and system
biology where structured theories from first principles as in physics
are basically absent.  However, in spite of some interesting attempts
in this direction, see e.g., \cite{Daniels2015}, we are still far from
this grand goal. Indeed some authors pointed out that the
reconstruction of dynamical equations from data is a computationally
NP-hard problem \cite{Cubitt}. 

It is however worth mentioning that some inference on the equation of
motion can be done when one has some ideas on the possible structure
of the equation of motions. For instance, for some specific systems,
like networks of nonlinearly coupled oscillators, with a record of the
phases evolution it is possible to use optimization methods with some
``educated guesses'' on the type of interactions (see,~e.g.,
Ref.~\cite{Piko1}), or using Bayesian inference
\cite{Stankovski2012}. Other interesting approaches, with a wider
scope, are based on so called symbolic regression \cite{Lipson}, in
which the functional form of the terms ruling the interactions
among the system variables is not completely preassigned. In a
nutshell the idea is to assume a set of building blocks (mathematical
structures). As the method requires a search in the
huge space of all possible interactions among the
variables, it is typically very slow and of limited
application.  However, we can mention recent improvements based on the
idea of sparseness, i.e., on the fact that only a few mathematical
structures are represented in certain classes of systems. In this case
using appropriate library of mathematical functions (e.g., polynomials
of two or three variables, or more elaborated functions) together with
powerful machine learning algorithms it has been possible to also
reconstruct the dynamics of complicated fluid dynamical systems
\cite{brunton}.  We notice that these methods are based on
reconstructing the r.h.s of Equations~(\ref{eq:map}) and (\ref{eq:ode}),
this means that the time record of the whole system state should be
sampled and with time intervals short enough to be able to reconstruct
the dynamics, i.e., the time derivative in the case of (\ref{eq:ode}).

A completely different perspective is to attempt predictions by using
virtually model-free methods. In this direction, besides a classical
approach---based on the so-called method of the analogues, used in many contexts, see e.g., \cite{loratm,lor69} for weather forecasting---that will be discussed in the next
subsection, which is applicable also when the whole state vector of the
system cannot be accessed, we briefly mention a recent
promising attempt based on machine learning techniques. 

In 2004 Jaeger and Haas \cite{Jaeger} proposed a technique known as
``reservoir computer''. In a nutshell the idea is to use the time
record of the system state, with dimension $D_{in}$, as input of a
neural network composed of three elements: An input layer with
precisely $D_{in}$ nodes (one for each of the system variables); a
nonlinear network with $D \gg D_{in}$ sparsely connected nodes (with
appropriate choice of the adjacency matrix of the internal network to
ensure some level of activity even in the absence of input, see
\cite{Jaeger} for details), which are then mapped into an output layer
with, again, $D_{in}$ nodes.  The network is then trained, using the
available time record, by requiring that the output reproduces the
input. The latter step is done leaving unchanged the adjacency matrix
of the internal network but modifying the weights that map the
internal $D$ nodes into the output $D_{in}$ nodes, with standard
techniques. Once the training is finished, the idea is to use a given
state of the system as input---the initial condition---and use the
output as new input. This way we have a ``new'' dynamical system that
evolves the state vector of the system of interest. In other words a
``model'' for the evolution which we do not fully control.  However,
as shown in Ref.~\cite{Jaeger}, the resulting dynamics predicts very
well the system evolution at least for low dimensional systems
(e.g., the Lorenz 1969 model). In the last years, the group of Ott has
shown that the same idea can be used to predict the evolution of
high-dimensional spatio-temporal chaotic systems
\cite{Ott1}. Strikingly, in Ref.~\cite{Ott1} it has been shown that
not only the ``new'' dynamical system can be used to predict the
evolution of the original system but it also provides the correct
spectrum of Lyapunov exponents, including the negative ones. Then some
extensions of the method have been proposed for very large systems
\cite{OttPRL}, where the idea is to use several internal networks for
different portions of the system, or to predict unmeasured variables
provided the training step was performed using the whole state vector
\cite{ottNew}.

\subsection{Prediction When We Do Not Know the System State Vector
\label{sub:analog}}
In typical situations we do not know the whole set of variables (not
even their number) defining the state of a system.  Moreover, even
knowing them, in experimental measurements, we usually have access
only to a few scalar observables $u(t)$ depending on the state:
$u(t)=G[{\bm x}(t)]$. In these cases, for not too pathologic function
$G$, there exists a powerful technique, based on Takens' \textit{delay
  embedding theorem} \cite{Takens1981,Sauer1991}, able to reconstruct
the phase space.  Essentially, these theorems establish that if the
system dynamics is effectively $D$-dimensional, where $D$ is not
necessarily the number of variables of the state space and is not
necessarily an integer number (see below), from a long data record of
a scalar observable, $u_k=u(k\Delta t)$ with $k=1,\ldots, M$ and
$\Delta t$ the sampling time, we can build a delay coordinate
reconstruction, i.e., define a vector in a $m-$dimensional space
\begin{equation}
\bm U_k^{(m)}=(u_k,u_{k-1},\ldots,u_{k-(m-1)})\,, 
\label{eq:delay}
\end{equation}
which is equivalent, in a sense specified below, to the original one
if $m\geq 2[D]+1$, where $[D]$ denotes the integer part of $D$.
Technically speaking, assuming the original dynamical system to be
dissipative, the dynamics will evolve onto a manifold, the attractor,
of Hausdorff (fractal) dimension $D<N$ (where $N$ is the
dimensionality of the state vector), which roughly represents the
number of ``active'' degrees of freedom of the system, below we will give
more specifications. Essentially the embedding
theorems ensure that the manifold reconstructed by the delay vectors
preserves the properties of the dynamical systems that do not change
under a smooth change of variables, such as, e.g., the Lyapunov~exponents.

The equivalence between the dynamics of the original state vector and
the delay coordinate is a very important result. Consider the simple
case in which the scalar series $\{u_{k}\}_{k=1}^M$ corresponds to the
evolution of one of the variables of the system. Then the embedding
theorem ensures that it suffices to reconstruct properties of the
whole set of variables composing the system. In particular, this means
that if we have access, e.g., to two variables, if they belong to the
same dynamical system (and thus they are causally dependent) the
delay reconstruction made using one of the variable is connected (by a
smooth change of coordinate system) to the reconstruction made using
the delay vector based on the other variable. For example, using this
property in Ref.~\cite{may2012} it was proposed a method to infer
causality, in a nutshell one searches for neighbors in one delay
reconstruction and study the correlation with neighbors in the
reconstruction based on the other variable.

Now let us discuss how one can use a time record to perform
predictions (we shall closely follow~\cite{cec12}).  In a nutshell the
basic idea, in its simplest formulation, can be summarized with the
old adage ``If a system behaves in a certain way, it will do it
again'', which finds its ground in the observation of regularities
(periodicity) such as, for instance, the diurnal and seasonal cycles.
This~idea, together with the belief in determinism ({\it from the same
  antecedents follow the same consequents}), it is at the basis of
prediction methods.  However, as Maxwell argued \cite{Maxwell}:
\textit{It is a metaphysical doctrine that from the same antecedents
  follow the same consequents.$[\ldots]$ But it is not of much use in
  a world like this, in which the same antecedents never again concur,
  and nothing ever happens twice.$[\ldots]$ The physical axiom which
  has a somewhat similar aspect is ``That from like antecedents follow
  like consequents.''}  These words warn us on the almost exceptional
character of periodic behaviors, indeed nowadays we know about the
ubiquitous presence of irregular evolutions due to deterministic
chaos.

A mathematical formalization of the above idea was due to Lorenz and it is
at the basis of the \textit{method of
 analogues} \cite{loratm,lor69}, which can be considered as
the most straightforward approach to predictability in the absence of
a detailed knowledge of the physical laws.

In its simplest implementation, the method works as follows. Assume
that the known state $\bm x(t)$ of a process can be sampled at times
$t_k=k\Delta t$ with arbitrary precision, for considerations on the
unavoidable presence of measurement noise we refer to the technical
literature on nonlinear time series analysis \cite{Kantz1997}. In the
case in which we only have a partial knowledge of the ``true'' state,
one can use delay vectors (\ref{eq:delay}). In the sequel, for the
sake of simplicity of presentation and without losing generality, we
can assume that that we have a time series of the system state,
sampled at each interval $\Delta t$, which is also assumed to be
arbitrary but not too short (see, e.g., \cite{Kantz1997} for a
discussion on the proper way to choose the sampling time in practical
cases). Given the sequence $\bm x_k=\bm x(t_k)$ with $k=1,\ldots M$,
how can we forecast the evolution of $\bm x_M$, i.e., predict $\bm
x_{M+T}$ at time $t_{M+T}$ ($T\geq 1$)? The basic idea is to search in
the past $(\bm x_1,\bm x_2,\ldots,\bm x_{M-1})$ that state, say $\bm
x_{k^*}$, most similar to $\bm x_M$, and to use its consequents as
proxies for the future evolution of $\bm x_M$. Mathematically, we
require that $|\bm x_{k^*} - \bm x_M|\leq \epsilon$, and we dub $\bm
x_{k^*}$ a $\epsilon$-\textit{analogue} to $\bm x_M$. We stress that when
  searching for analogues in the delay reconstructed space one should
  be aware that if the embedding dimension is not appropriately
  chosen, a lot of \textit{false} analogues may be present, for a
  discussion on how to test and avoid this problem see, e.g.,
  \cite{Kantz1997}. If the analogue were perfect ($\epsilon=0$) the
system (being deterministic) would be surely periodic and the
prediction trivial: $\bm x_{M+T}\equiv \bm x_{{k^*}+T}$ for any $T$.
If it were not perfect ($\epsilon >0$), we could use the forecasting
recipe  
\begin{equation}
\hat{\bm x}_{M+T}= \bm x_{{k^*}+T}\,,
\label{eq:simplepred}
\end{equation}
as \textit{from like antecedents follow like consequents}.  For the
prediction (\ref{eq:simplepred}) to be meaningful, the analogue $\bm
x_{k^*}$ must not be a near-in-time antecedent.  A straightforward generalization of
(\ref{eq:simplepred}) is thus to use as a prediction the average over
the future evolution of all $\epsilon$-analogues of $\bm x_M$
\cite{Kantz1997}.

Once a ``good'' analogue, i.e., with $\epsilon$ reasonably small, has
been found, the next step is to determine the accuracy of the
prediction. With reference to Equation~(\ref{eq:simplepred}) this means to
quantify the evolution of the error $|\hat{\bm x}_{M+T}-\bm x_{M+T}|$.
In practice, the $\epsilon$-analogue is the present state with an
uncertainty, $\bm x_k=\bm x_M+\delta_0$ ($\delta_0\leq
\epsilon$). Therefore, if the underlying system is chaotic one expects
the error to grow exponentially and thus we go back to
Equation~(\ref{eq:predtime}). So that, apparently, the main limit to
predictions based on analogues is the sensitive dependence on initial
conditions, typical of chaos.  But, as realized by Lorenz himself, the
main issue is actually to find good (small $\epsilon$) analogues
\cite{lor69}: \textit{In practice this procedure may be expected
  to fail, because of the high probability that no truly good
  analogues will be found within the recorded history of the
  atmosphere.}  He also pointed out that, even in the presence of
chaos, the main issue with the method relies on the need of a very
large data set \cite{loratm}.

In the rest of this section we will illustrate why it is so difficult to
find ``good'' (small $\epsilon$) analogues. We shall begin recalling
some basic notions and facts from ergodic theory of dynamical systems
that will then be illustrated with a numerical example.

The founding idea of ergodic theory is that the long-time statistical
properties of a system can be equivalently described in terms of the
invariant (time-independent) probability, $\mu$, such that $\mu(S)$ is
the probability of finding the system in any specified region $S$ of
its phase space.  If the trajectories of a $N$-dimensional ergodic
system evolve in a bounded phase space $\Omega\in \mathbb{R}^N$, the
Poincar\'e recurrence theorem \cite{Poincare} ensures that analogues
exist as it proves that the trajectories exiting from a generic set
$S\in \Omega$ will return back to such set $S$ infinitely many
times. The theorem holds for almost all points in $S$ except for a
possible subset of zero probability. It was originally formulated for
Hamiltonian systems, but it can be straightforwardly extended to
dissipative ergodic systems provided initial conditions are chosen on
the attractor, i.e., on the set $A\in \mathbb{R}^N$ onto which the
system states asymptotically evolve, and ``zero probability'' is
interpreted with respect to the invariant probability, $d\mu(\bm x)$,
on the attractor. In $N$-dimensional dissipative systems, the
attractor $A$ has typically a dimension $D_A<N$. Slightly more
formally, the dimension $D_A$ describes the small scale ($\epsilon \ll
1$) scaling behavior of the probability $\mu \bigl(B^N_{\bm
  y}(\epsilon) \bigr)$ of finding points $\bm x\in A$ which are in the
$N$-dimensional sphere, $B_{\bm y}(\epsilon)$, of radius $\epsilon$
around $\bm y$:
\begin{equation}
\label{frac-dim}
\mu \bigl ( B^N_{\bm y}(\epsilon) \bigr ) = \int_{B^N_{\bf y}(\epsilon)} d\mu({\bm
  x}) \sim \epsilon^{D_A} \,.
\end{equation}
when $D_A$ is not integer, the attractor and the probability measure
on it are dubbed fractal. Typically, $D_A$ depends on the point $\bm
y$ (i.e., the measure on the attractor is
multifractal~\cite{libro}). For the sake of simplicity, we shall
ignore this in the following and assume that there is a single
dimension $D_A$, i.e., we assume the attractor to be homogeneous.

Poincar\'e theorem merely proves that a trajectory (almost) surely returns to
the neighborhood of its starting point---i.e., the existence of
analogues---, but does not provide information about the time between
two consecutive recurrences---the Poincar\'e recurrence
time. The latter is however
crucial to the method of analogues. Indeed, to hope to find an
$\epsilon$-analogue of given state one needs the time record of the
system state to have a duration of the order of the recurrence time.
Another result in ergodic theory, the Kac's lemma \cite{kac},
is about the recurrence time. It states that given a set $S$
including ${\bm x}_0$ (i.e., the state we are interested in), for an ergodic system, the mean recurrence time
of $\bm x_0$ relative to $S$, $\langle \tau_S(\bm x_0) \rangle$, is inversely
proportional to the measure of the set $S$ \cite{kac}, i.e.,
\begin{equation}
\label{kac1}
\langle \tau_{S}(\bm x_0) \rangle \propto \frac{1}{\mu(S)} \,,
\end{equation}
where the average is computed over all the states $\bm x\in S$
according to the invariant measure. Now~taking $S=B^N_{\bm
  x_0}(\epsilon)$ and using (\ref{frac-dim}) yields
\begin{equation}
\label{rtimeDa}
\langle \tau_{\epsilon}(\bm x_0) \rangle \sim \epsilon^{-D_A} \,.
\end{equation}

Therefore, if we require $\epsilon$ to be very small and it happens
that $D_A$ is large the average recurrence time becomes indeed huge.

As the above description is somehow formal, it is useful to illustrate it 
with a  specific example. We consider the following nonlinearly coupled
ordinary differential equations,  first proposed by Lorenz in 1996
\cite{Lorenz1996}:
\begin{equation}
\label{l96}
\frac{ \textrm{d} X_n} {\textrm{d} t} \!=\! 
X_{n-1}(X_{n+1}\!-\!X_{n-2}) -X_{n}+ F\, , \qquad n\!=\!1,\ldots, N\,,
\end{equation}
with periodic boundary conditions ($X_{N\pm n}=X_{\pm n}$).  The
variables $X_n$ may be thought of as the values of some atmospheric
representative observable along the latitude circle, so that
Equation~(\ref{l96}) can be regarded as a one-dimensional caricature of
atmospheric motion \cite{Lorenz1996}. The quadratic coupling conserves
energy, $\sum_n X^2_n$. In the presence of forcing $F$ and damping
$-X_n$, the energy is only statistically conserved. The motion is thus
confined to a bounded region of $\mathbb{R}^N$. Moreover, dissipation
constrains the trajectories to evolve onto a subset---the attractor---of this region possibly with dimension $D_A<N$. The dynamical
features are completely determined by the forcing strength $F$ and the
system dimensionality $N$. For~$F > 8/9$ and $N \ge 4$ the system is
chaotic \cite{lore2scale}.

Assuming ergodicity, Poincar\'e theorem  holds with respect to the invariant measure and for any
generic state we should expect analogues to exist for any
$\epsilon$. Let us now count them.  Being interested in typical
behaviors and not just in the properties around a specific state we
select $r$ states $\{\bm X_\star^{k}\}^r_{k=1}$ along a given
trajectory of (\ref{l96}), well spaced in time to sample
the attractor, and we compute the
average fraction of their $\epsilon$-analogues as
\begin{equation}
\label{grasspro}
C_{r,M}(\epsilon)= \frac{1} {Mr} \sum_{k=1}^{r}\sum_{j=1}^M 
\Theta( \epsilon -\vert{\bm X}_j- {\bm X}_\star^{(k)}\vert)  \, .
\end{equation}
 
Practically, as we know the evolution laws (\ref{l96}), instead of
looking at the backward time series of the reference states, we can
select the $\{\bm X_\star^{k}\}^r_{k=1}$ and look at their forward
$\epsilon$-analogues.

Now we notice that the average fraction of $\epsilon$-analogues
Equation~(\ref{grasspro}) is nothing but the fraction of time the
trajectory spends in a sphere of radius $\epsilon$ centered around the
reference states. Thus, for large $M$, as a consequence of ergodicity,
$C_{r,M}(\epsilon)$ is simply the average probability to visit these
$\epsilon$ balls around the reference points.  Therefore, with the
assumption that only the dimension $D_A$ characterizes the system, for
sufficiently large $M$ and small $\epsilon$, Equation~(\ref{frac-dim})
implies  $C_{r, M} (\epsilon) \sim
\epsilon^{D_A}$. Actually, the quantity $C_{r,M}(\epsilon)$ is nothing
  but an approximate computation of that so called correlation
  integral introduced by Grassberger and Procaccia~\cite{Grassberger1983} to
  compute the correlation dimension, $D_2$, of strange attractors
  emerging from dissipative chaotic systems. Here the assumption of
  homogeneity implies $D_2\approx D_A$. We notice that accounting for
  the heterogeneity is a mere technical complication which does not
  change the main points of this section.

\begin{figure}
  \includegraphics[clip=true,width=0.5\textwidth]{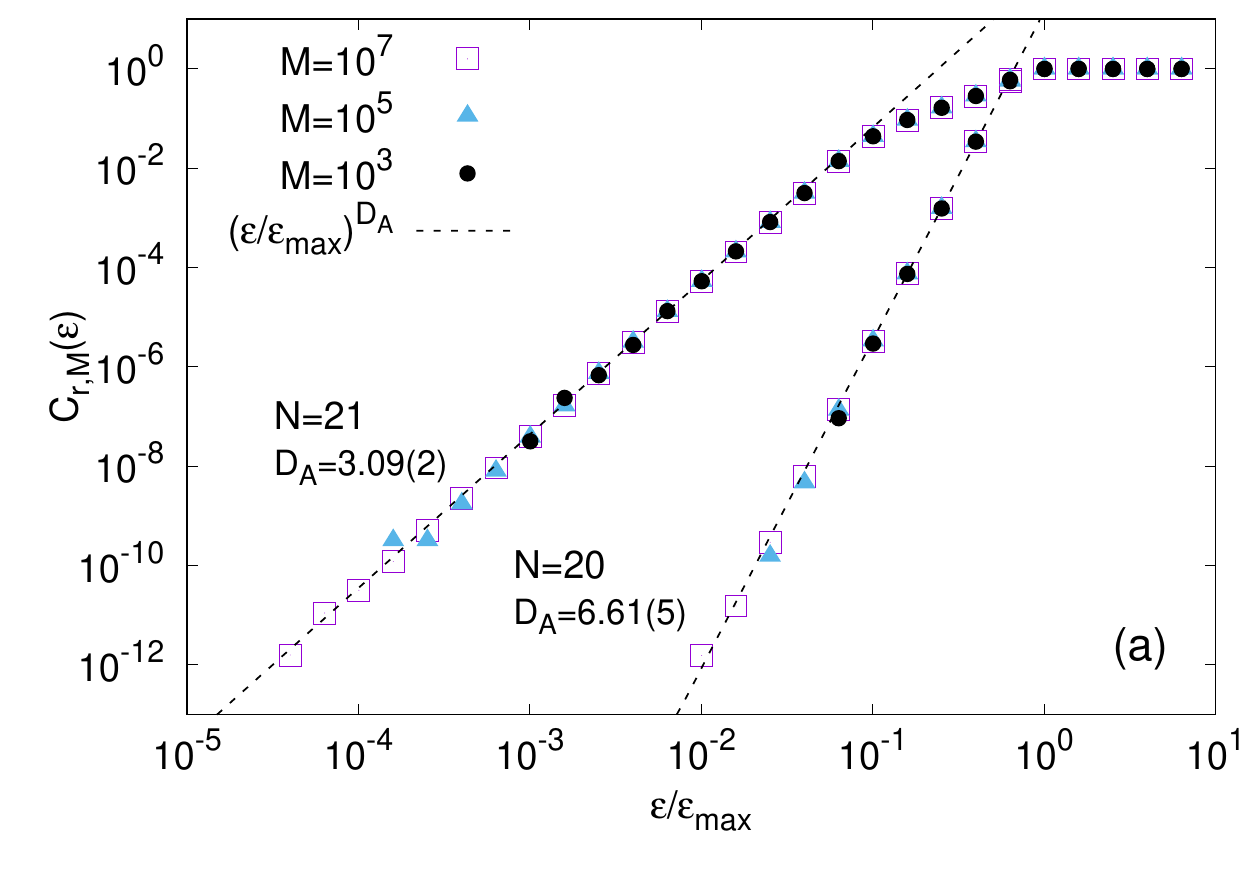}
  \includegraphics[clip=true,width=0.5\textwidth]{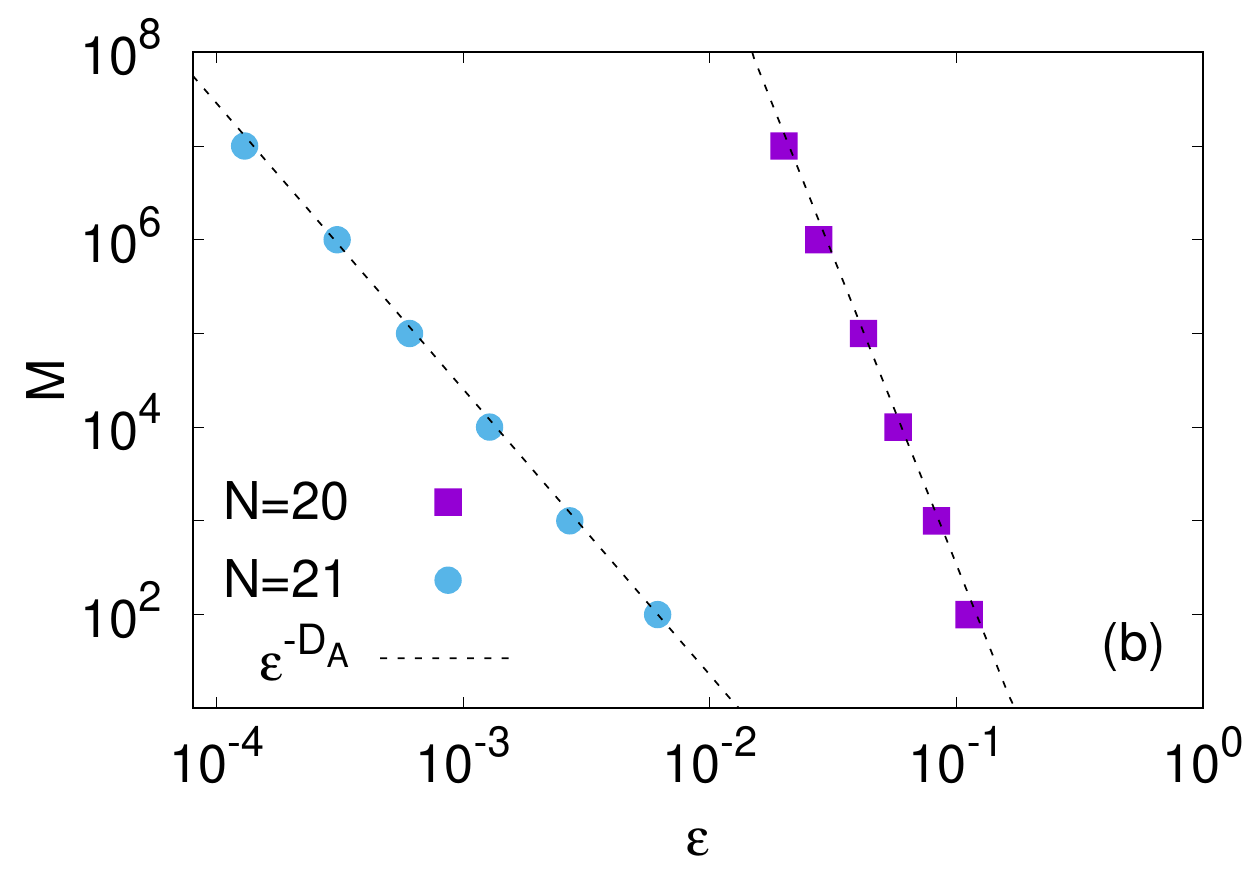}
\caption{ (\textbf{a}) $C_{{r},M}(\epsilon)$
  vs. $\epsilon/\epsilon_{\textrm{max}}$ for $F=5$, $N=20$ and $N=21$;
  the reference states are ${r}=1000$ and different values of $M$
  ranging from $10^3$ to $10^7$ are considered. The solid lines are
  the fits of the data assuming $C_{r, M} (\epsilon) \sim
  \epsilon^{D_A}$; (\textbf{b}) Number of data points $M$ vs
  $\epsilon_{min}/\epsilon_{max}$ vs. $M$. The dashed lines are the
  fits of the data by means of relation~(\ref{M-df}).\label{corrX}}.
\end{figure}

In Figure~\ref{corrX} we show $C_{r,M}(\epsilon)$ obtained with
${r}=10^{3}$ reference states and different lengths $M$ of the time
series, from $10^3$ to $10^7$.  When the degree of similarity
$\epsilon$ becomes larger than the attractor size, say
$\epsilon_{\mathrm{max}}$, the fraction $C_{r,M}(\epsilon)$ saturates
to $1$. Therefore, we normalize the degree of
similarity by $\epsilon_{\textrm{max}}$.  As for the dynamics
(\ref{l96}), the forcing is fixed to $F=5$ and we consider two system
sizes $N=20$ and $N=21$. In both cases the system is chaotic.  The
solid lines in Figure~\ref{corrX} indicate that for $\epsilon \ll
\epsilon_{\textrm{max}}$ the probability to find an analogue is fairly
well approximated by the expected power law $C_{r, M} (\epsilon) \sim
\epsilon^{D_A}$. Fitting the data with this expression, we find
$D_A\simeq 3.1$ and $D_A\simeq 6.6$ for $N=21$ and $N=20$,
respectively. Practically, this difference in $D_A$ means that
 the probability to find $\epsilon$-analogues with
$N=20$ becomes about $\epsilon^{3.5}$ times smaller than with $N=21$.

Slightly changing the viewpoint, we observe that $C_{r,M}(\epsilon)$
relates to the Poincar\'e recurrence times we mentioned above. To
understand the latter point, fix $r=1$, i.e., consider the analogues to
the state $\bm x_M$ in a sequence $\bm x_1, \ldots, \bm x_{M}$ sampled
at each $\Delta t$ and denote their number with
$\mathcal{M}(\epsilon)$.  Clearly, the average time interval, $\overline{\tau_\epsilon}(\bm x_M)$, between two consecutive $\epsilon$-analogues of $\bm
x_M$ is
\begin{equation}
\overline{\tau_\epsilon}(\bm x_M) = \frac{(M-1)\Delta t}{\mathcal{M}(\epsilon)} = \frac{\Delta t}{C_{1,M}(\epsilon)} \propto \epsilon^{-D_A}\,,
  \label{eq:kac}
\end{equation}
where we used that $C_{1,M}(\epsilon)=\mathcal{M}(\epsilon)/(M-1)$,
being it the fraction of $\epsilon$-analogues. Thus
$C_{r,M}(\epsilon)$ is simply a further average of such times over
different balls of radius $\epsilon$ centered on different reference
states. Clearly, in order to find an $\epsilon$-analogue we must
require the number of points $M$ to satisfy $M\Delta t \geq
\overline{\tau_\epsilon}(\bm x_M)$, that from (\ref{eq:kac}) implies $M\geq
1/C_{1,M}(\epsilon)$ and, given the hypothesis of homogeneity of the
attractor we have $C_{1,M}(\epsilon) \sim C_{r,M}(\epsilon) \sim
\epsilon^{-D_A}$, we can realize that the minimum length of the time
series is
\begin{equation}
\label{M-df}
M \sim \Bigl(\frac{L}{\epsilon} \Bigr)^{D_A} \,\, ,
\end{equation}
$L$ being the typical excursion of each component of 
${\bm  x}$, used to nondimensionalize the expression.

Equation~(\ref{M-df}) implies that, at least in principle, the method
can work for deterministic systems having an attractor of finite
dimension provided the time series is suitably long. However, the
exponential dependence on $D_A$ in Equation~(\ref{M-df}) imposes, upon
putting the numbers, too severe constraints. For instance in
Figure~\ref{corrX}b, we show how the number of data-points $M$ scales
with distance between a reference point and its best analogue
($\epsilon_{\mathrm{min}}$).  We see that for
$\epsilon_{\mathrm{min}}/\epsilon_{\mathrm{max}}=10^{-2}$ a sequence
of $10^2$ points is sufficiently long in the case $N=21$ ($D_A \approx
3.1$) while, on the contrary, even $10^7$ points are not yet enough in
the case $N=20$ ($D_A \approx 6.6$), the rationale for this difference
is Equation~(\ref{M-df}).  We notice that the counter-intuitive inequality
$D_A(N=21)<D_A(N=20)$ is a peculiar detail of the considered model
\cite{lore2scale}. Generally, $D_A$ is expected to increase with $N$
\cite{libro}. The values of $N$ and $F$ here used are motivated to
emphasize the importance of the effective number of degrees of freedom
that, in general, is not trivially related to (and can be much smaller
than) the number of variables $N$.

We stress that the necessity of an exponentially large (with $D_A$)
amount of data constitutes a genuine intrinsic difficulty of every
analysis based on time series without any guess on the underlying
dynamics.  Such a difficulty is not a peculiarity of the method of
analogues, but is inherent to all methods based on the occurrence
frequency of sequences of states to estimate the average of
observables.  The~problem arises whenever one needs to collect enough
recurrences, including the Grassberger and
Procaccia \cite{Grassberger1983} technique or the method to infer
causality \cite{may2012}.  Though these considerations may sound trivial,
it is worth recalling that, in the '80s, when nonlinear time series
analysis started to be massively employed in experimental data
analysis, the limitations due the length of the time series were
overlooked  and a number of misleading papers
appeared even in important journals~\cite{Ruelle1989}.

Summarizing, in virtue of (\ref{M-df}), when $D_A\gtrsim 5-7$ the
number of data-points required for finding good analogues is
prohibitive. This sounds very pessimistic for predictions from data,
it is worth concluding with some words of optimism. Indeed, the
problem of finding the analogues may be mitigated in the presence of
a multiscale structure, where the vector state ${\bm x}$
can be decomposed into a slow component ${\bm X}$ which is also the
``largest'' one, and a fast component ${\bm y}$ ``small'' with respect
to ${\bm X}$ (i.e., $y_{\mathrm{rms}} \ll X_{\mathrm{rms}}$). In such
cases, provided that the slow components can be described in terms of
an ``effective number'' of degrees of freedom much smaller than those
necessary to characterize the whole dynamics, mediocre (referred to
the whole system) analogues can be used to forecast at least the
slower evolving component.

We illustrate this point by considering a variant of  (\ref{l96}) introduced by Lorenz himself \cite{Lorenz1996} to discuss
the predictability problem in the atmosphere, where indeed a
multiscale structure is present. The model reads
\begin{eqnarray}
\strut{\hspace{-0.6truecm}}\frac{ \textrm{d} X_n} {\textrm{d} t} \!&=&\!  
X_{n-1}(X_{n+1}\!-\!X_{n-2})\!-\!X_n\!+\!F \!-\! \frac{hc}{b}
 \sum_{k=1}^{K} y_{k,n}    \label{lore2scaleA}\\
\strut{\hspace{-0.6truecm}}\frac{ \textrm{d} y_{k,n}} {\textrm{d} t} \!&=&\! 
cb\,  y_{k+1,n}(y_{k-1,n}\!-\!y_{k+2,n})\!-\!c\,y_{k,n} +\frac{hc}{b}  X_n,
    \label{lore2scaleB}
  \end{eqnarray}
with $n=1,\ldots,N$ , $k=1,\ldots,K$ and boundary conditions $X_{N \pm
  n}=X_{\pm n}$, $y_{K+1,n}=y_{1,n+1}$ and
$y_{0,n}=y_{K,n-1}$. Equation~(\ref{lore2scaleA}) is the same as
(\ref{l96}) but for the last term which couples $\bm X$ to $\bm
y$. The~variables $\bm y$ evolve with a similar dynamics but are $c$
times faster and $b$ times smaller in amplitude. The parameter $h$,
set to $1$, controls the coupling strength. In geophysical fluid
dynamics the variable $\bm X$ and $\bm y$ represent the synoptic and
convective scales, respectively \cite{Lorenz1996}.

\begin{figure}
\centering
\includegraphics[clip=true,width=0.7\columnwidth]{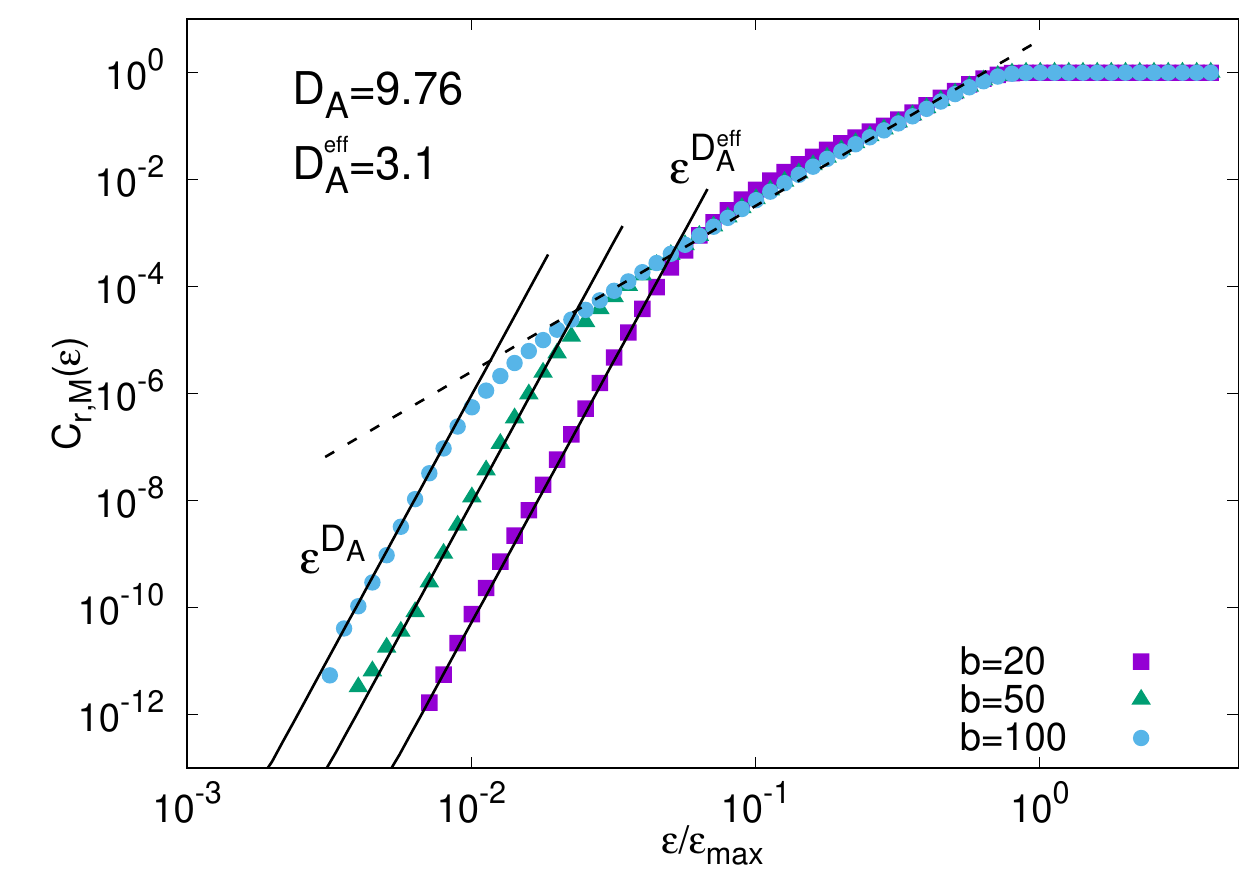}
\caption{ $C_{r,M}(\epsilon)$ vs. $\epsilon/\epsilon_{max}$ for model
  (\ref{lore2scaleA}--\ref{lore2scaleB}) computed for three scale
  separations $b$ (as labeled) holding the other parameters fixed at
  $h=1$, $c=10$, $F=10$, $N=5$ and $K=10$. The dashed line has slope
  $\approx 3.1$ while the solid lines have all the same slope $\approx
  9.76$.  The quantity (\ref{grasspro}) has been computed with
  $r=10^3$ and $M=10^7$.\label{multi} }
\end{figure}

Figure~\ref{multi} shows the fraction of $\epsilon$-analogues
$C_{r,M}(\epsilon)$ for the system
(\ref{lore2scaleA}--\ref{lore2scaleB}) obtained by assuming that
a long record, $M=10^7$, of the whole state of the system $\bm x(t)=(\bm X(t),\bm y(t))$ can be measured.  In this example the time scale
separation is fixed to $c=10$ and the fast component $\bm y$ is
$b=20, 50$, and $100$ times smaller than the slow one $\bm X$. The
phase-space dimensionality is $55$, with $N=5$ slow and $K=10$ fast
degrees of freedom for each slow one. The attractor dimension of the
whole system $D_A$ is rather large ($D_A
\approx 10$).  However, for $\epsilon/\epsilon_{\mathrm{max}} >
O(1/b)$ a second power law, $C(\epsilon) \sim
\epsilon^{D_A^{\mathrm{eff}}}$ with $D_A^{\mathrm{eff}} \approx 3 <
D_A$, appears defining a sort of ``effective dimension at large scale''.
Therefore, if we are interested in predicting the slow evolving
component of the system, provided it is described by a relatively low
number of effective degrees of freedom, as here, we can exploit the
mediocre analogues (i.e., the $\epsilon$-analogues with
$\epsilon/\epsilon_{\mathrm{max}} > O(1/b)$). Moreover, it is
reasonable to expect that the prediction error related to mediocre
analogues grows in time as $ \sim \epsilon e^{\lambda(\epsilon) t}$ where
$\lambda(\epsilon)$ can be much smaller than the Lyapunov exponent
$\lambda_1$ (indeed as shown in Ref.~\cite{BOFFETTAETAL}
$\lambda(\epsilon) \approx \lambda_1/c$).  This~implies that slow
variables can be predicted over longer term than the whole state of
the system, as already realized by Lorenz \cite{Lorenz1996}.  We
remark that the above example is a very simplified idealization of a
multiscale system and something much more complicated may happen, see
e.g., \cite{Olbrich}.

We conclude this Section by noticing that unlike the Lorenz 1963
(\ref{B4}), the 1996 models (\ref{l96}) and
(\ref{lore2scaleA}--\ref{lore2scaleB}) were not obtained by means of an
approximation of hydrodynamic equations, but have been built just
using the type (quadratic) of nonlinearities and the conservation laws
(i.e., of energy in the inviscid limit) of the original system.  They
represent toy models which preserve crucial aspects of the original
problem with the advantage of simplifying the numerical approach. For instance,
Equations~(\ref{lore2scaleA})--(\ref{lore2scaleB}) is a playground to
understand the dynamics of systems at very different time and spatial
scales. In this class of models we mention the shell models for
turbulence which have been built to ``mimic'' the Navier-Stokes
equations and allowed for understanding (even quantitatively) many
aspects of the energy cascade in turbulence
\cite{bohr2005dynamical,libro}.

\section{Protein Modeling: An Example of Data-Driven Approach\label{sec:protein}}
The impressive progress undergone by Molecular Biology in the last
decades produced a wealth of knowledge and accurate data around
protein molecules.  All this body of information needs to be
rationalized into theoretical frameworks enabling interpretation and
prediction of elementary biological processes. In this context, the
study of proteins well illustrates how models can be built upon and
improved by using experimental data to address more challenging tasks.

Proteins are biopolymers performing or assisting a myriad of
biological processes; to become active they need to assume a specific
three-dimensional structure: The native or active state (NS). Under~physiological conditions proteins fold spontaneously into the NS in a
way that is mainly encoded by the sequence of aminoacids
\cite{Anfinsen}, and the accurate prediction of the structure from the
sequence is a longstanding issue of Molecular Biology known as {\em
  protein folding problem} \cite{FinkelBook}.  Computational methods
based on clever modeling are the primary tools to assist and
complement experimental research in solving the problem.

Before illustrating the key elements of protein models, it is necessary to understand
which is the basic feature of a good model. This arises from the comparison 
between natural proteins and random heteropolymers \cite{FinkelBook}, in fact  
proteins are known to be anything but random objects.

\begin{figure}
\centering
\includegraphics[clip=true,keepaspectratio,width=10.0truecm]{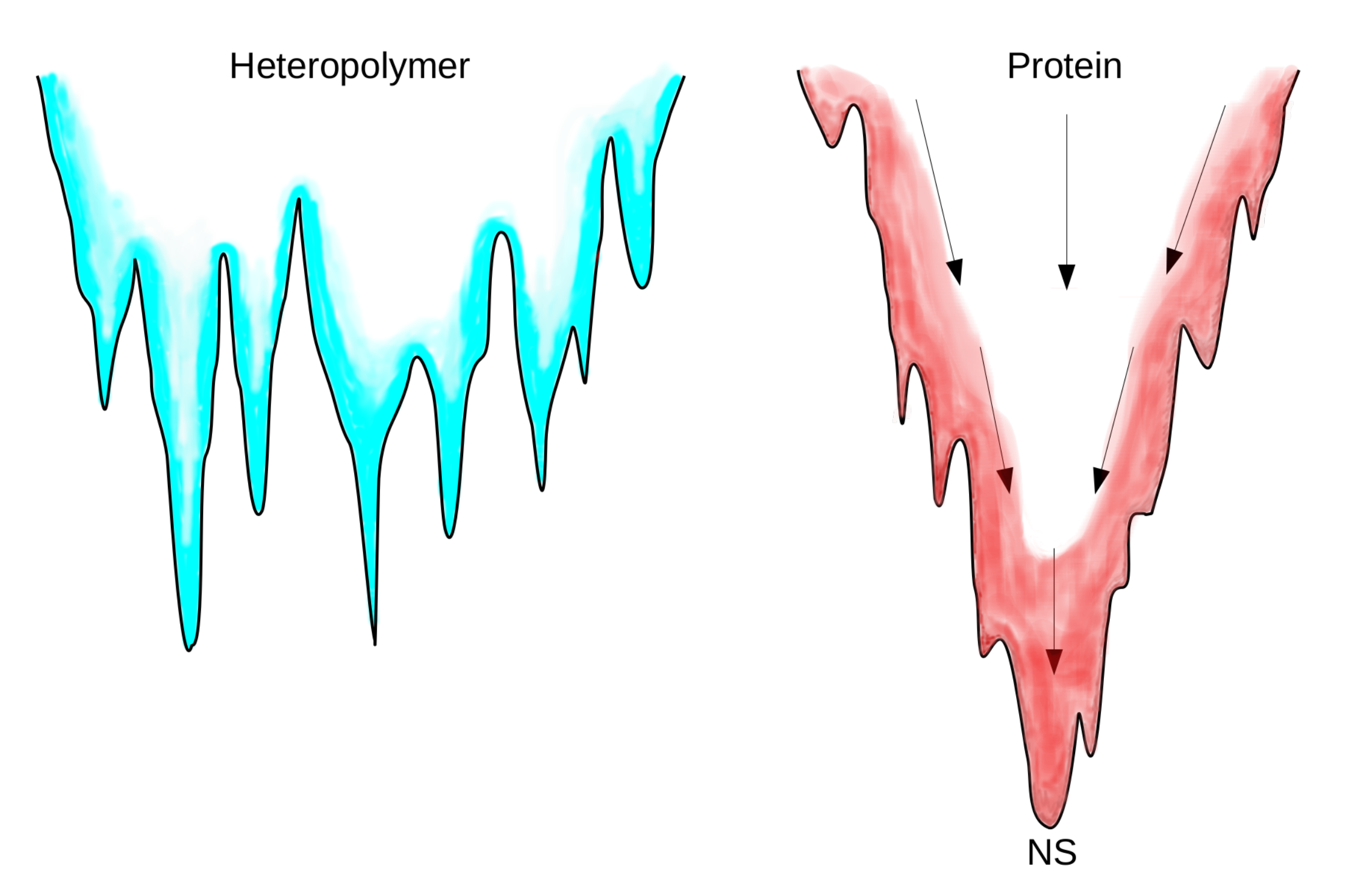}
\caption{Sketch of the energy landscapes: Glassy-like for random
  heteropolymers (\textbf{left}) with several deep minima and high barriers and
  funnel-like for proteins (\textbf{right}) with the native state (NS) occupying the
  ``unique''~minimum.}
\label{fig:funnel} 
\end{figure}

The difference between proteins and heteropolymers clearly emerges
from the structure of their respective energy
landscape~\cite{Onuchic,Plotkin}, see Figure \ref{fig:funnel}.  The
energy landscape of random heteropolymers is studded
with many degenerate minima, each one representing the end-point of
the folding process.  The trapping action of such deep metastable
minima slows down the folding process.  In contrast, protein-like
landscapes have been ``sculpted'' by evolution in a funnel-like shape,
where NS is placed at the bottom, Figure~\ref{fig:funnel}.  The protein,
following the funnel, experiences a simultaneous decrease of entropy
and energy so that no high barriers emerge along the pathway to the
NS.  Therefore, the funnel confers thermodynamic stability and fast
accessibility to the NS.

The landscape picture strongly suggests that protein models should reproduce a 
funnel-shaped energy landscape with a moderate amount of frustration. 

A general expression of the potential energy of classical
physics-based all-atom model consists of various interaction
terms~\cite{Schlick}
\begin{equation}
U = U_{bond} + U_{angle} + U_{dihe} + U_{vdw} + U_{elec} 
\label{eq:allAt}
\end{equation}

\begin{figure}
\centering
\includegraphics[clip=true,keepaspectratio,width=8.0truecm]{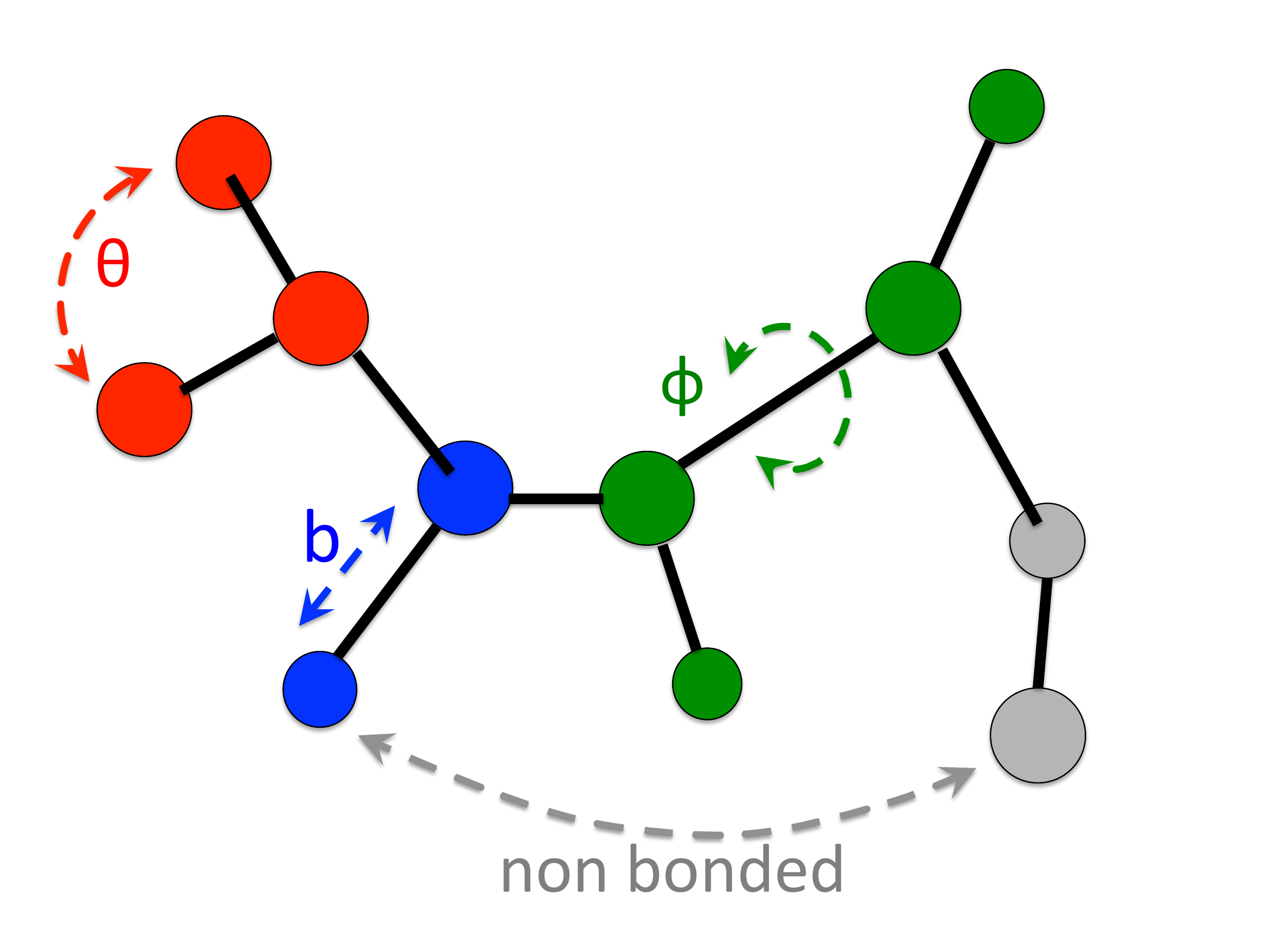} 
\caption{\label{fig:intera} Typical interactions considered in a
  all-atomistic force field. Atoms of a molecules interact by simple
  springs, in lengths and angles (bonded contributions), and via long
  range forces (non bonded contributions).  In a mechanical view the
  forces determine the spatial arrangements of atoms an their~dynamics.}
\end{figure}

The first three of them, so-called “bonded” terms, describe the
principal structural deformations (sketched in Figure \ref{fig:intera})
due to bond stretching (blue), bending (red), and torsion for a
rotation about certain dihedral angles (green).  The last two terms,
referred to as “non-bonded” interactions, describe dispersion and
repulsion effects (Lennard-Jones term) and electrostatic interactions.

\unskip

Despite its conceptual simplicity, the energy function
\eqref{eq:allAt} presents two big practical problems when applied to
the simulation of real systems.  First, for proteins of medium and
large size, it involves many atoms of different species with a
undesired proliferation of terms and parameters.  Second, both
realistic parameters and units need to be assigned.  This step,
generally referred to as the {\em parameterization of the
  force-field}, is rather delicate if one requires maximal
applicability of the model (transferability).  A discussion of these
issues is beyond our purpose, here we limit ourself to mention the
general strategy.  Parameters are optimized to match the increasing
body of experimental information: molecular geometries, thermodynamics
data and other spectroscopic data.  In particular, vibrational spectra
of small molecules, detected by infrared radiation and Raman
scattering \cite{spectro}, allow for obtaining the force constants
and inter-atomic distances of bonded interactions (terms 1,2,3 in
Equation \eqref{eq:allAt}).  If the measured vibrational frequency of a
specific bond is $\omega$, in the model, the strength of that
interaction will be set to $k \simeq \omega^2 M$, with $M$ being the
reduced mass of the atoms involved.

Non-bonded interactions can be obtained by the Boltzmann-Inversion
method \cite{Chandler}. Suppose, for instance that a scattering
experiment sampled the pair-correlation function $g_{ij}(r)$ between
atoms $i$ and $j$, then it is natural to assign them the interaction
potential,
$$
u_{ij}(r) = C - k_B T \ln[g_{ij}(r)]\,,
$$
where the potential is known up to an additive constant $C$.
The above formula stems from the {\em reversible work theorem}
\cite{Chandler}, a general result of
Statistical Mechanics proving that the pair correlation function of an
equilibrium system satisfies the relation
$$
g_{ij}(r) = \exp\{-\beta w_{ij}(r)\}
$$ where $w_{ij}(r)$ is the reversible work (free energy) spent to
bring the tagged particles $i,j$ from an infinite separation to a
distance $r$.
This inversion formula can be also applied recursively 
from an initial guess of potential to improve the potential 
estimate \cite{Schommers}.

In practice, we are in the position to know many elementary
rules driving a chain of aminoacids into the native structure, but our
algorithmic implementation of the model \eqref{eq:allAt} is currently
so inefficient that only reproduces folding events of small protein
molecules.  Real folding processes of larger structures are still
prohibitive to all-atom simulations.

Therefore, a clever strategy of simplifications is needed which brings
to the building of Coarse grained (CG) protein models
\cite{Tozzini_OP,Clementi_OP,guardCG,Kolinski}.  A CG procedure
basically tries to keep a similar form of the all-atom force field
\eqref{eq:allAt} while operating a drastic decimation of degrees of
freedom to reduce the computational load at the price of losing
resolution.  In this respect, coarse graining means that groups of
atoms are joined together to form ``pseudo-atoms'' (also named
``united-atoms'') which work as centers of interaction in the new
simplified model.  This procedure, which to preserve the phenomenology
must be not too invasive, requires re-definition and
re-parameterization of ranges and intensities of the interactions
among pseudo-atoms.

The definition of the CG-model interactions follows the same strategy
discussed for the all-atomistic force fields, the simplification
arises from the reduced number of parameters, however the conceptual
difficulty remains.  For CG-models there is the additional advantage
that the parameterization can also be performed by using the results
of the full-atom simulations. This~can be addressed by the powerful
techniques of inverse-problems and statistical inference, such as
max-entropy and maximum likelihood principles \cite{inverse}.

Below we illustrate two CG-approaches that use only 
protein structural information to derive reasonable interactions 
among aminoacids. In this respect they constitute examples of models 
uniquely driven by the empirical knowledge.      

\subsection{Knowledge Based Approach}
The accuracy of experimental techniques allows 
protein databases to be updated at an impressive rate with new experimentally 
determined structures.
This rich information can be exploited to ``learn'' empirical interactions 
among aminoacids, via a statistical analysis that looks 
for recurrent folding rules over a large number of known protein structures. 

Tanaka and Scheraga \cite{knowledge} suggested to use a coarse-grained
potential obtained from the frequency of aminoacid contacts observed
in a set of crystal structures of proteins.  A contact means that two
aminoacids are neighbors in a given structure, and this proximity is
interpreted as the presence of an interaction.  Accordingly, the
interaction energy between two types of amino acids, $a,b$, can be
assigned~as
$$
\epsilon_{ab} = -k_B T \ln\bigg[\dfrac{N_{ab}}{N_{ab}(0)}\bigg]
$$
where $N_{ab}$ is the observed frequency of the contact of the 
couple $a-b$ in the representative database. 
While $N_{ab}(0)$ denotes the expected frequency of that contact in a chosen 
reference state, that is a set of disordered (i.e., non-native) conformations 
where that contact is only ``randomly'' established. 
Therefore meaningful interactions emerge from the comparison between 
the probability of finding them in native-like structures 
and the probability of occurrence in the reference state of non-native 
conformations. 
Of course, the quality of the results depends on the ``good choice'' of 
both database and reference state, for this reason, their optimal selection 
is the subject of debate and much investigation.

The knowledge-based interaction parameters are rather general as they 
not only incorporate many effects acting between atoms, 
such as electrostatic, van der Waals forces, etc.,  
but also the influence of the surrounding solvent.

It is interesting to remark that the principle for generating 
empirical potentials is equivalent to the methods of analogues
discussed in Section~\ref{sub:analog}. 
In the former, recurrent regularities 
are searched in the database of structures, while the latter 
searches for recurrences in the time series of events.  

\subsection{Structure Based Models}
The concept of funnel-like landscape (Figure~\ref{fig:funnel}) is
carried to extremes in the structure based modeling, originally
introduced by G\={o} and coworkers~\cite{Ueda75}.  These models are
built up in such a way that energy function attains its minimum on the
coordinates of the crystallographic structure of the NS.  Therefore
the experimental data enter crucially in the model definition.  A
simple way to achieve that NS is a minimum is by introducing the
notion of native interactions, or {\em contacts}. Two atoms are said
to form a native contact if in the NS they are neighbors, i.e., if
their native distance $R_{ij}<R_c$ for a chosen $R_c$.  We assume that
only atoms forming native contacts interact attractively, whereas the
others repel each other via an excluded-volume potential.  As in this
framework a decrease in energy can only be achieved through an
increase in the fraction of native contacts, the G\={o}-model
minimizes frustration, assuring a fast and correct arrival to the NS.

A popular G\={o}-like force field was introduced by Clementi et al. \cite{Clementi} 
who proposed for a $N$-beads system the potential-energy: 
\begin{eqnarray*}
\Phi_{G\bar{o}} & = &\sum_{i=1}^{N-1} \frac{k_{p}}{2}(r_{i,i+1}-R_{i,i+1})^{2} + 
\sum_{i=1}^{N-2} \frac{k_{\theta}}{2}(\theta_{i}-\Theta_{i})^{2} + \\ \nonumber 
& & \sum_{i=1}^{N-3} \big\{k_{\varphi}^{(1)}[1 - \cos(\varphi_{i} - \Phi_{i})] +
k_{\varphi}^{(3)}[1 - \cos (3\varphi_{i} - 3\Phi_{i})]\big\} 
+ V_{nb}(r_{ij}) \,.
\end{eqnarray*}

The first term, that enforces chain connectivity, 
is a stiff harmonic potential between consecutive residues allowing only small 
fluctuations of the bond lengths $r_{i,i+1}$  around their native values $R_{i,i+1}$.
Likewise, the elastic potential in $\theta$ allows only small 
fluctuations of the bending angles $\theta_{i}$ around their native values, $\Theta_{i}$.
The native-like secondary structure (helices, beta-sheets) is 
enforced by a potential in the dihedral angles. Each $\varphi_i$ 
angle, identified by four consecutive beads, characterizes the local 
torsion of the chain. Again,  $\Phi_{i}$ denotes the value of the $i$-th angle in 
the native structure.
Finally, the long-range potential $V_{nb}$, which favors the formation of the 
correct native tertiary structure by promoting attractive interactions,
is the two-body function
\begin{eqnarray*}
V_{nb}(r_{ij}) = \epsilon 
\begin{cases} 
5\left(\dfrac{R_{ij}}{r_{ij}}\right)^{12} - 
6\left(\dfrac{R_{ij}}{r_{ij}}\right)^{10} & R_{ij} \leq R_c\\[10pt]
      \dfrac{10}{3}\;\left(\dfrac{\sigma}{r_{ij}}\right)^{12} & R_{ij} > R_c\;.
\end{cases}
\end{eqnarray*} 

Therefore, the interaction between aminoacids $i-j$ is attractive when
their distance in the native structure, $R_{ij}$, is below a certain
cutoff, $R_{c}$, otherwise the aminoacids repel each other via a
soft-core $\sigma$.  It means that the system gains energy as much as
its conformation is close to the NS.  A unique parameter $\epsilon$
sets the energy scale of the force field and the
  other parameters introduced above are the typical ones used in
similar G\={o}-type approaches, see e.g., \cite{Clementi,cecconi2006testing}.

This philosophy may seem artificial, as the folding problem is somehow
reversed, because the sequence is neglected, and the NS becomes not
the final goal but the root of the model.  However, a number of
empirical evidences supports the use of G\={o}-models, by underscoring
the importance of the topology of the NS, which is encoded by the
adjacency matrix representing the network of native contacts.  In
particular, proteins with different sequences exhibit similar folding
mechanisms if they have similar native and transition states.
Moreover, simple topological parameters characterizing the complexity
of the NS were found to correlate well with the folding rates of
small globular proteins~\cite{Plaxco98,Chiti99}, again showing the
relevance of the native structure.

G\={o} models can be successfully used to sample the Transition State
Ensemble, which is the collection of protein conformations that have
the same probability (=1/2) either to refold in the NS or to unfold.
This knowledge allows the computation of folding rates and the
understanding of folding mechanisms for globular
proteins~\cite{Hills08,GuardianiC5,cecconi2006testing,Clementi}.  Moreover,
since G\={o}-like models by construction well describe the NS
geometry, they offer the opportunity to study the stability of
structures during the unfolding~\cite{Kleiner,Li2007}.

The main criticism raised to the G\={o}-like philosophy is the
principle of minimal frustration, which leads to the absence of
non-native interactions.  Indeed, studies pointed out the relevance of
energetic frustration in real proteins because non-native contacts may
form in the transition state without being present in the
NS~\cite{Paci}. However, the frustration due to non-native contacts is
not always deleterious, but seems to have certain relevance in
favoring the folding process.
At first glance, the introduction of increasingly larger non-native
energies is expected to slow down the folding process, however Plotkin
and Clementi \cite{Plotkin04} reached the counter intuitive conclusion
that a moderate amount of frustration reduces the free energy barriers
and increases folding rates.

Another criticism of G\={o} models concerns the exclusion of sequence
effects. In G\={o} models the folding is driven only by the topology
of the NS, but it is the sequence that determines the topology. The
precise role of the sequence in folding remains to be understood.  A
common strategy to account for sequence effects is by using
heterogeneous contact energies that may be chosen using different
criteria again based on structural information \cite{Karanicolas,
  Dokholyan}.

We conclude this brief survey on CG protein modeling with the final
consideration ``whether CG models will stand the test of time'' as
they may appear just as computational tricks to by-pass the
limitations of current computational resources.  This idea is
misleading, as the ability of these models to correctly reproduce some
experimental patterns shows that not all the molecular degrees of
freedom are equally important. In other words, the CG approaches are
able to single out the relevant driving forces among the multitude of
chemical details of macromolecules.  This is why CG models will remain
precious conceptual tools for the study of macromolecular systems even
when the advances in computer science will make atomistic simulations
feasible on biologically relevant timescales.

\section{Langevin Equation: When a Multiscale Structure Helps \label{sec:langevin}}
Several problems in science are characterized by the presence of very
different time scales. Among the most important examples, we mention
protein folding and climate: for proteins, the time scale of the
vibration of covalent bonds is $O(10^{-15})$ s, while the folding time
can be of the order of seconds; in the case of climate, the
characteristic times of the involved processes vary from days (for the
atmospheric phenomena) to $O(10^3)$ yr for the deep ocean flows and
ice shields.  The necessity of treating the ``slow dynamics'' in terms
of effective equations is both practical (even modern supercomputers
are not able to simulate all the relevant scales involved in certain
difficult problems) and conceptual: Effective equations are able to
catch some general features and to reveal dominant ingredients which
can remain hidden in the detailed description. The study of such
multiscale problems has a long history in science, and some very
general mathematical methods have been
developed~\cite{castiglione2008, engquist, givon}, whose usage,
however, is often not easy in non-idealized situations.

Historically, the theory of Brownian motion represents perhaps the
first example of multiscale physical modeling. In their studies
Einstein, Smoluchowski, and Langevin recognized the multiscale
structure intrinsic in the movement of a particle of $\sim$1 $\mu$m
size suspended in a fluid made of molecules of $\sim$1 $\AA{}$ size,
and exploited such a structure to achieve an effective theoretical
model. The Brownian motion model (simple diffusion) deduced by
Einstein and Smoluchowski has become the first brick for the future
development of the theory of stochastic
processes~\cite{vanKampen2007,gardiner}. With a different spirit but
similar phenomenological arguments (the Stokes law), and statistical
assumptions (thermal equilibrium of the colloidal particle with the
liquid), Langevin introduced his celebrated stochastic differential
equation~\cite{langevin,lemons} which---after more than a century---stands as an unrivaled paradigm for small systems displaying smooth
slow dynamics superimposed with observable fast fluctuations. Most
importantly, a fundamental modern wave of non equilibrium statistical
mechanics, that is stochastic thermodynamics, is founded upon the
Langevin model~\cite{vanKampen2007,Seifert,livipoliti}.

A natural question is the possibility to derive the Langevin Equation
(LE) in a non phenomenological way, i.e.,  starting from the dynamics
of large systems \cite{zwanzig}.  Unfortunately, there are just few
cases where it is possible to use such a desirable
approach~\cite{castiglione2008}. One situation is the motion of a
heavy particle in a diluted gas: In such a case, using an approach
going back to Smoluchowski, with a statistical analysis of the
collisions among the heavy particle and the light gas particles, it is
possible to determine the viscous
friction~\cite{von1906kinetischen,cecconi2007transport}. A complete
derivation, including the shape of the noise term, has been also
obtained as a perturbative expansion of the Lorentz-Boltzmann equation
by van Kampen~\cite{vK61}, repeated in a similar form for granular
gases~\cite{sarracino}. A mathematical account of this procedure can
be found in~\cite{lebowitz}. There is also another large class of systems where it is possible to
obtain a LE in an analytical way: Harmonic chains with a heavy
particle of mass $M$ and $N$ light particles of mass $m \ll
M$~\cite{rubin60,turner60,zwanzig73}.  In such a case, the linearity of
the dynamics allows for an explicit solution and then the possibility
to find the LE for the heavy particles in the limit $m/M \ll 1$ and $N
\gg 1$.

As far as we know there are no other clearly distinct cases where it
is possible to derive a LE starting from a deterministic mechanical
model of the whole system. It is certainly desirable to have the
possibility to write down a LE for a heavy particle also in systems
different from the well known cases above discussed. Approximate
derivations for the case of non-linear forces~\cite{maes17} and for
cases near non-equilibrium stationary processes~\cite{maes15} have
also been recently considered. A particularly interesting test-case is
that of systems where the Hamiltonian has a non-standard kinetic term,
i.e., non-quadratic in the momentum, leading to non-trivial properties
such as negative absolute
temperature~\cite{cpv,tempreview,termometri,baldolang}, a possibility
recently verified also in experiments~\cite{Braun2013}.

In the following we discuss a very different approach, that is a
practical procedure to build a Langevin equation from a long time
series of data from experimental or numerical results. Up to
our knowledge a similar procedure has been applied, previously, to
many model systems~\cite{peinke2} but only recently some of us have
applied it to Hamiltonian systems, including systems displaying
negative temperatures~\cite{baldolang}. The procedure discussed here
does not include a preliminary check for the validity of a LE
description, it blindly assumes such a validity: Of course one then
has to wonder in a critical way if such an assumption is correct, or
at least, consistent.  In the large category of multiscale phenomena,
there are different physical ingredients which can provide a
justification for a LE description. In many cases---in analogy with
what discussed in the above examples---the necessary separation of
time scales is guaranteed by the existence of a massive (slow) degree
of freedom with respect to the ``background'', i.e., the aforementioned
condition $M \gg m$. Possible exceptions where such a separation of
masses can be insufficient for a LE is discussed
in~\cite{baldolang}. In real situations one simply assumes the
existence of two classes of variables (slow and fast) with well
separated time scales and tries to reconstruct a LE for the slow variables. A
posteriori, the quality of such an initial assumption can be verified
by comparing the original data with the results of the
model~\cite{peinke,peinke2,kantz,engquist,givon}. Some authors have
discussed a general mechanism justifying the scale-separation
assumption, within a dynamical systems approach~\cite{kifer,mackay}.

We briefly illustrate the LE reconstruction protocol. The basic
assumption for this procedure is that the slow variable,
say ``$v$'', can be described by a Ito-Langevin equation with the shape
\begin{equation} \label{langevin}
\frac{d v}{d t}=F(v) + \sqrt{2D(v)} \eta(t)
\end{equation}
where $\eta$ is a white noise with $\langle \eta(t)\rangle=0$ and
$\langle \eta(t) \eta(t') \rangle =\delta (t- t')$.  The function
$F(v)$, as well as $D(v)$, can be obtained from a long time series of
the values of $v(t)$ at time intervals $\Delta t$.  Being
$$
\Delta v(t)\equiv v(t+\Delta t) - v(t)
$$
we have for consistency with~\eqref{langevin} that
\begin{subequations}\label{coeff}
\begin{align}
F(v)&=\lim_{\Delta t \to 0} \frac{1}{\Delta t} \langle \Delta v(t) | v(t)=v \rangle\\
D(v)&=\lim_{\Delta t \to 0} \frac{1}{2\Delta t} \langle \left[\Delta v(t)-\Delta t F(v)\right]^2 | v(t)=v \rangle \,.
\end{align}
\end{subequations}

Assuming stationarity, the quantities on the r.h.s. do not depend on
time and they can be estimated as temporal averages, for several
choices of $\Delta t$. By extrapolating their dependence on the time
interval in the limit $\Delta t \rightarrow 0$, one can infer a good
approximation for the drift and the diffusivity.

Of course the practical evaluation of such limits is not completely
straightforward, since it involves a careful analysis of the relevant
time-scales of the process. In particular, the approximation of the
dynamics of a deterministic process as a LE cannot be valid at any
arbitrarily small time-scale, but only for $\Delta t$ larger than a
certain threshold (sometimes called the ``Markov-Einstein
time''~\cite{peinke2}): the limit $\Delta t \to 0$, therefore, must be
interpreted in a proper physical way.

In several contexts the method has been applied also to Markovian
processes in which the independent variable is a spatial coordinate,
instead of time $t$: This is the case, for instance, of the
reconstruction of rough surfaces performed in Ref.~\cite{jafari03}.
In that work the height $h(x)$ of a copper film deposited on a
polished Si(100) substrate at a certain time has been studied as a
function of the spatial coordinate $x$.  As a first step, the
markovianity of the stochastic process $h(x)$ has been carefully
checked; then, the described method has been performed considering a
long data series and the reconstructed surface has been compared to
the original one, finding a good agreement. Another class of processes
which can be treated with this method is that of quantities that
depend on the observed temporal (or spatial) scale, e.g., in turbulence
analysis \cite{renner01}. Once the Markov property of the process has
been verified, the procedure can be exploited in a completely
analogous way.

In principle a slight modification of the method could also be applied
to discontinuous stochastic processes, and a generalized LE with Lévy
noise could be inferred \cite{siegert01}. In these cases the first
step is the determination of the Lévy stability parameter $\alpha \in
(0,2]$, which can be measured from long time-series by exploiting the
  relation:
\begin{equation}
 \ln \langle \left(| \Delta v(t)-\Delta t F(v)|\right)| v(t)=v \rangle
 = \frac{1}{\alpha}\ln \Delta t +C(\alpha)\,.
\end{equation}

One can evaluate the l.h.s. as a temporal average for several values
of $\Delta t$ and then infer $\alpha$ as the inverse of the slope.

Further examples are discussed in some detail in Ref.~\cite{peinke2},
where a complete illustration of the method is given and its
widespread application fields are overviewed.

To conclude we give a brief demonstration of the power of the
LE-reconstruction procedure for three systems whose common feature is
the existence of a slow degree of freedom. For the three cases we have
a long time series of the velocity of the slow particle and, assuming
that it can be fairly approximated by a Markov process, we infer the
parameters of a LE.  Characteristic properties of the reconstructed
model---such as the stationary probability density function, the
autocorrelation or power spectrum, and the mean squared displacement---are then computed analytically or by numerical simulations, and the
results are compared with the original data.

{\em Harmonic chain} The first system we consider is a chain of $2N+1$ coupled harmonic oscillators with Hamiltonian
\begin{equation} \label{ham_osc}
 H=\frac{P^2}{2M} + \sum_{i=\pm 1, ..., \pm N} \frac{p_i^2}{2m} +
 \frac{k}{2}\sum_{i=-N}^{N+1} (q_i-q_{i-1})^2, \quad \quad
 q_{-N-1}\equiv q_{N+1}\equiv 0
\end{equation}
 in which the heavy particle, referred to as the
 \textit{intruder}, occupies the central position ($Q\equiv q_0$). The~parameter $k$
 represents the elastic constant, while $m$ and $M$ are the mass
 of the light particles and that of the intruder, respectively. We
 adopt fixed boundary conditions for the first and the last particles
 for computational reasons: they prevent an unbounded drift of
 positions caused by the conservation of total momentum.
 
 This system can be solved analytically. In a slightly modified version it
 has been extensively studied since 1960, when Rubin and Turner, in
 their seminal works \cite{rubin60,turner60}, showed that the behavior
 of the heavy particle could be approximated by a Brownian motion,
 under the assumption of canonically distributed initial
 conditions. In particular, for the autocorrelation function of the
 heavy particle's velocity $C(t)$ one has
 \begin{equation}
  C(t) \sim \exp\left(-\frac{2\sqrt{k m}}{M-m} t \right) + O\left(\frac{m}{M}\right)
 \end{equation}
 when $M/m \gg 1$. Further analyses \cite{takeno-hori62} on the
 frequency spectrum of the normal modes pointed out that the previous
 approximation was valid only if the ratio $M/\sqrt{k m}$ continued to
 be finite when the heavy mass limit was taken. Several
 generalizations of this simple model have been explored: The linear
 chain with nearest-neighbors interactions has been shown to be just a
 particular case of a wider class of harmonic systems with similar
 properties \cite{mazur-braun64,ford-kac-mazur65}, that can be
 used as ``thermal baths'' for the intruder even if the heavy particle
 is subjected to non-linear forces \cite{zwanzig73}. Further details
 on the numerical protocol and its application in this specific case
 can be found in~\cite{baldolang}.

\begin{figure}
 \centering
 \includegraphics[width=0.49\linewidth]{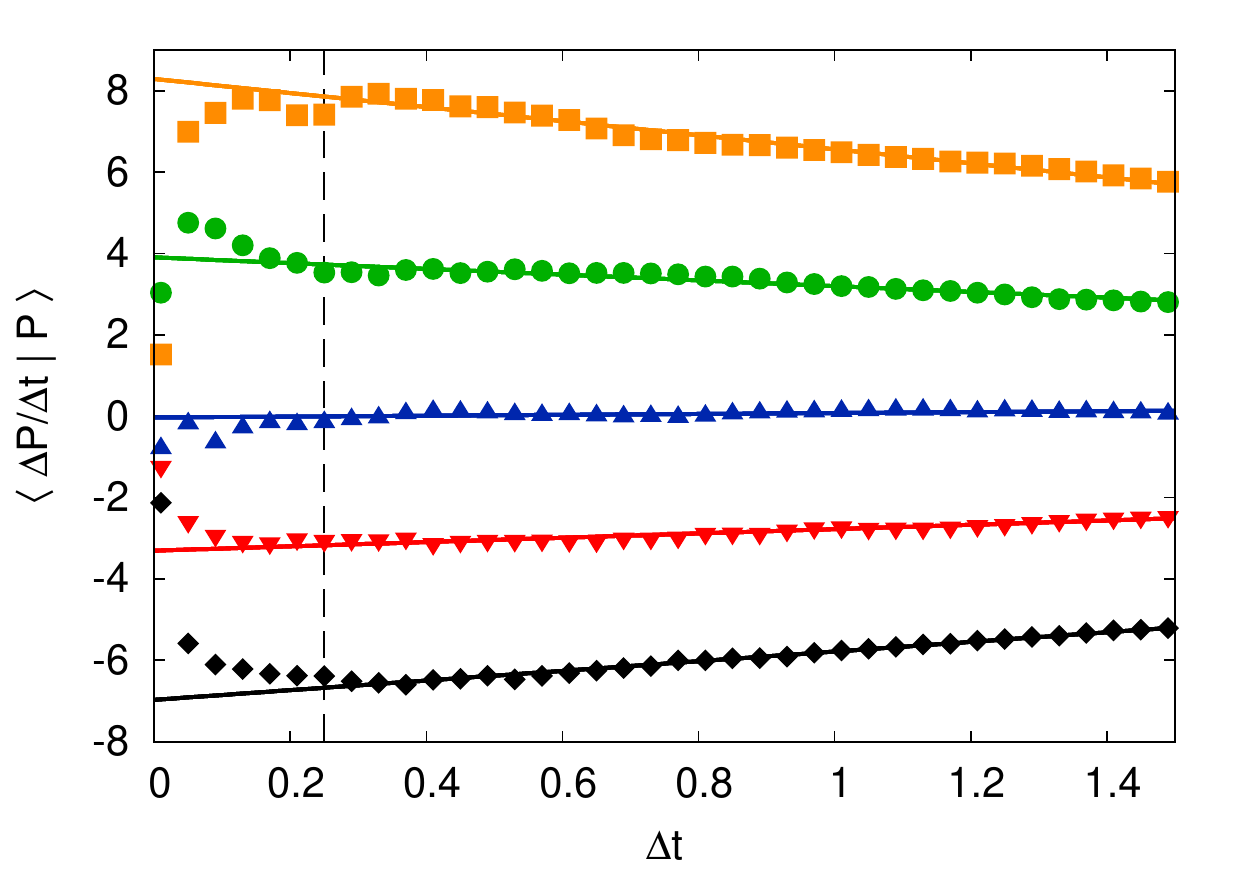}
 \includegraphics[width=0.49\linewidth]{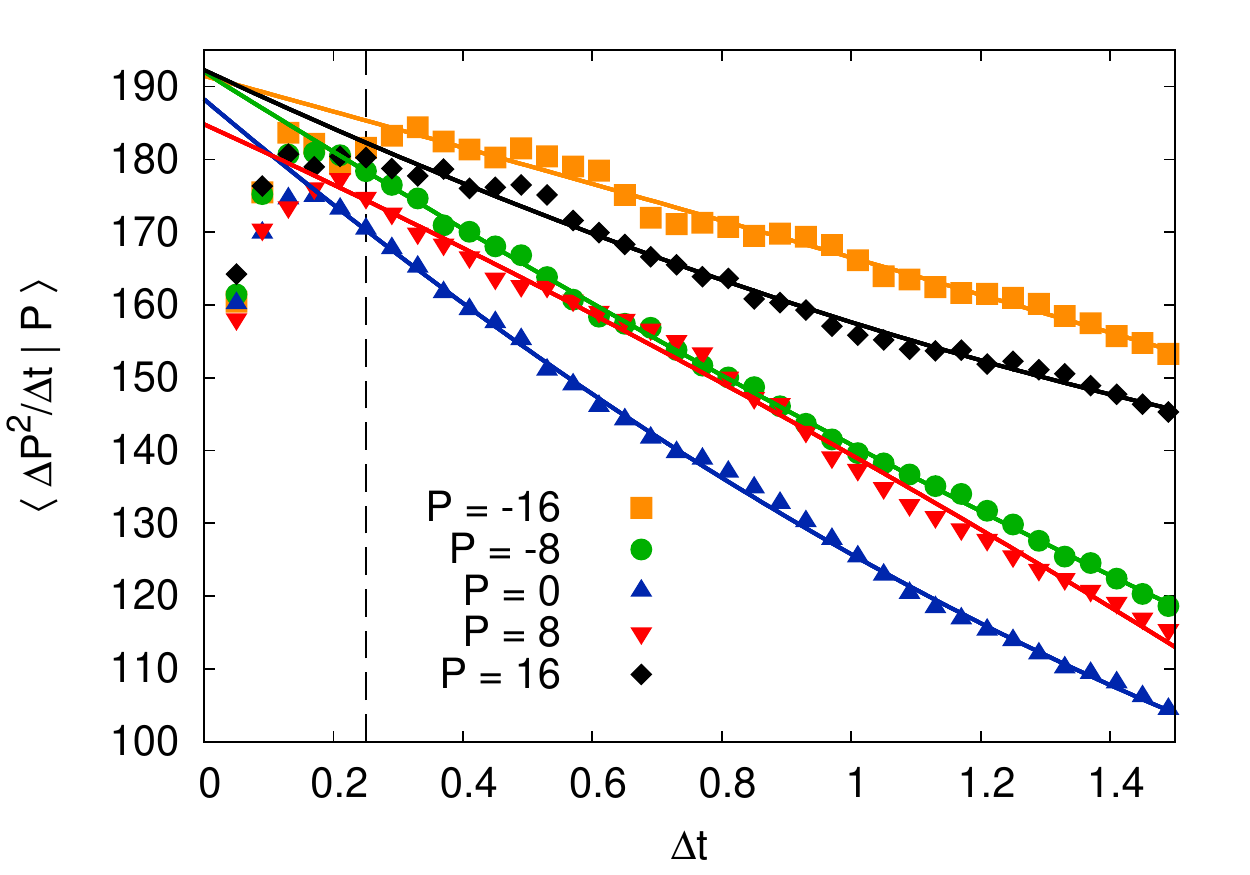}
 \includegraphics[width=0.49\linewidth]{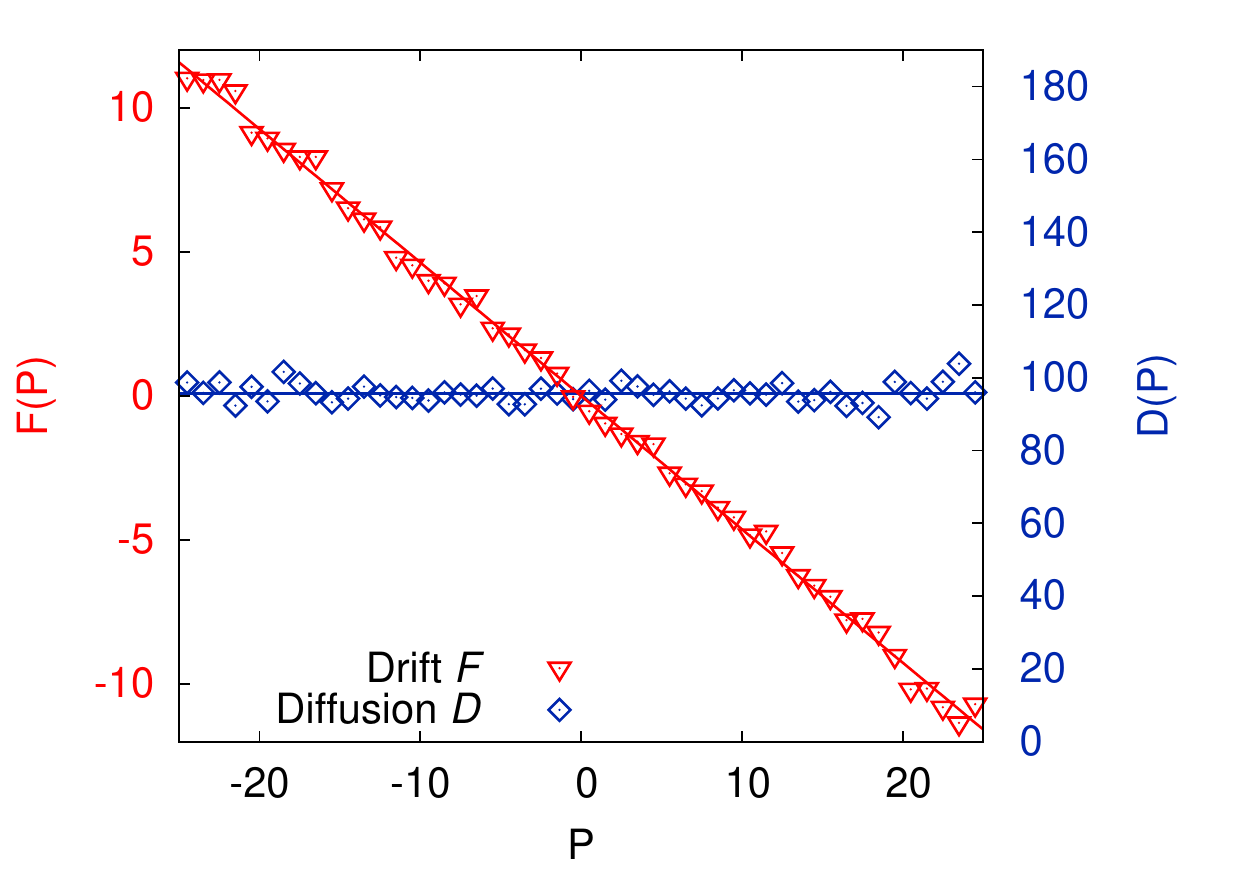}
 \includegraphics[width=0.49\linewidth]{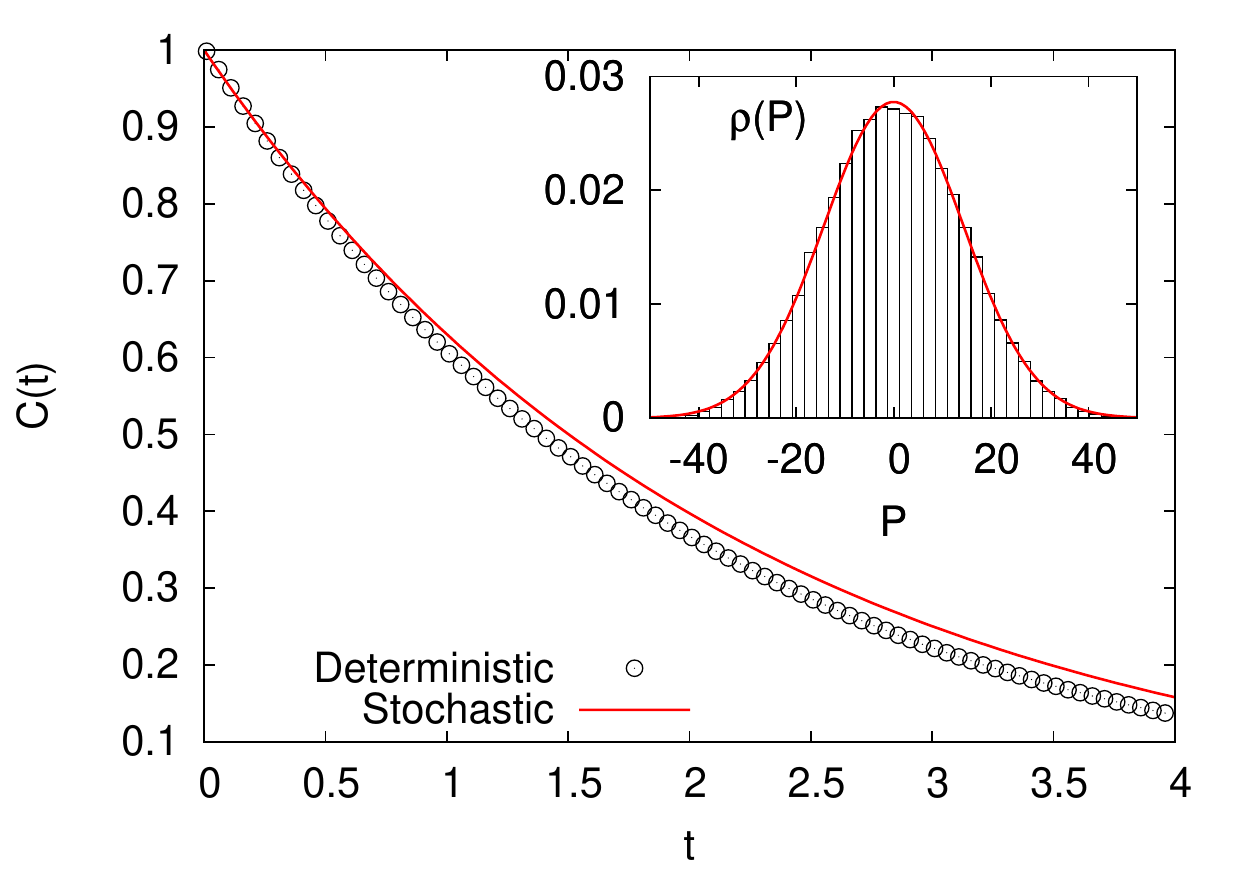}
 \caption{Harmonic chain. The model is Equation~\eqref{ham_osc} with
   $M=200$, $k=2500$, $2N=2000$, $\beta\simeq1.0$; integration step
   $\delta t=10^{-3}$.  Top: Extrapolation of the limits \eqref{coeff}
   in order to estimate drift (\textbf{left}) and diffusivity (\textbf{right}), for
   several values of $P$.  Bottom left: Drift and diffusion
   coefficients for the Langevin Equation describing $P$, inferred
   from simulations. Bottom right: Autocorrelation function for the
   velocity $\dot{Q}$ of the intruder; black circles represent the
   outcomes of molecular dynamics simulations, solid red line is
   computed by simulating LE with the previously inferred
   coefficients. Inset: Momentum p.d.f.  from the same deterministic
   (histogram) and stochastic (solid red line) simulations.}
 \label{fig:harmonichain}
 \end{figure}

 The procedure is perfectly successful in reconstructing the
 parameters of the Langevin Equation from the data, as can be seen in
 Figure~\ref{fig:harmonichain}.

{\em Elastic gas} As second model we consider a $2D$ gas of elastic  hard
  disks with a very low fraction of occupied volume. We simulate
  $N=1000$ disks: Of these $999$ have mass $m=1$ and $1$ has mass
  $M=100$ (the ``tracer''). All the disks have diameter
  $d=2r=0.01$. They move with periodic boundary conditions in a square
  of area $A=L\times L=10$. The number density is $n=N/A=100$, the
  coverage (or ``packing'') fraction is $\phi=\pi*N*r^2/A=0.785 \%$,
  i.e., the system is very dilute. We initialize the system with total
  kinetic energy $K=1000$ which is exactly conserved by the
  dynamics. This implies a kinetic temperature $T=\langle v_{i}^2 \rangle/2=M \langle v_0^2 \rangle/2=1$ with $i
  \neq 0$.
  
  For a massive tracer in a dilute gas the dynamics of the Cartesian
  components of the tracer's velocity $v_0 \equiv (V_x,V_y)$ follow
  two independent linear LEs i.e., Equation~\eqref{langevin} with
  $F(v)=-\gamma v$ and $v=V_x$ or $v=V_y$.  The damping coefficient is
  given by (see for
  instance~\cite{lebowitz,cecconi2007transport,sarracino})
  \begin{equation}
\gamma = 2\sqrt{2\pi} n d\sqrt{\frac{T}{m}}\frac{m}{M}=\frac{2\sqrt{2\pi}}{100} \approx 0.05.
\end{equation}

Finally the noise amplitude is determined by the Fluctuation-Dissipation theorem $D=\gamma T/M$~\cite{ma}.

Performing the extrapolation procedure, we find that the reconstructed
damping force can be described, as expected, as a linear function of
the velocity, while the diffusivity is almost constant. Upon comparing
the inferred parameters to the theoretical predictions discussed
above, one finds a very good agreement (relative difference $\sim$1\%
for $\gamma$, $\sim$5\% for $\sqrt{2D}$). 

\begin{figure}
 \centering
 \includegraphics[width=0.49\linewidth]{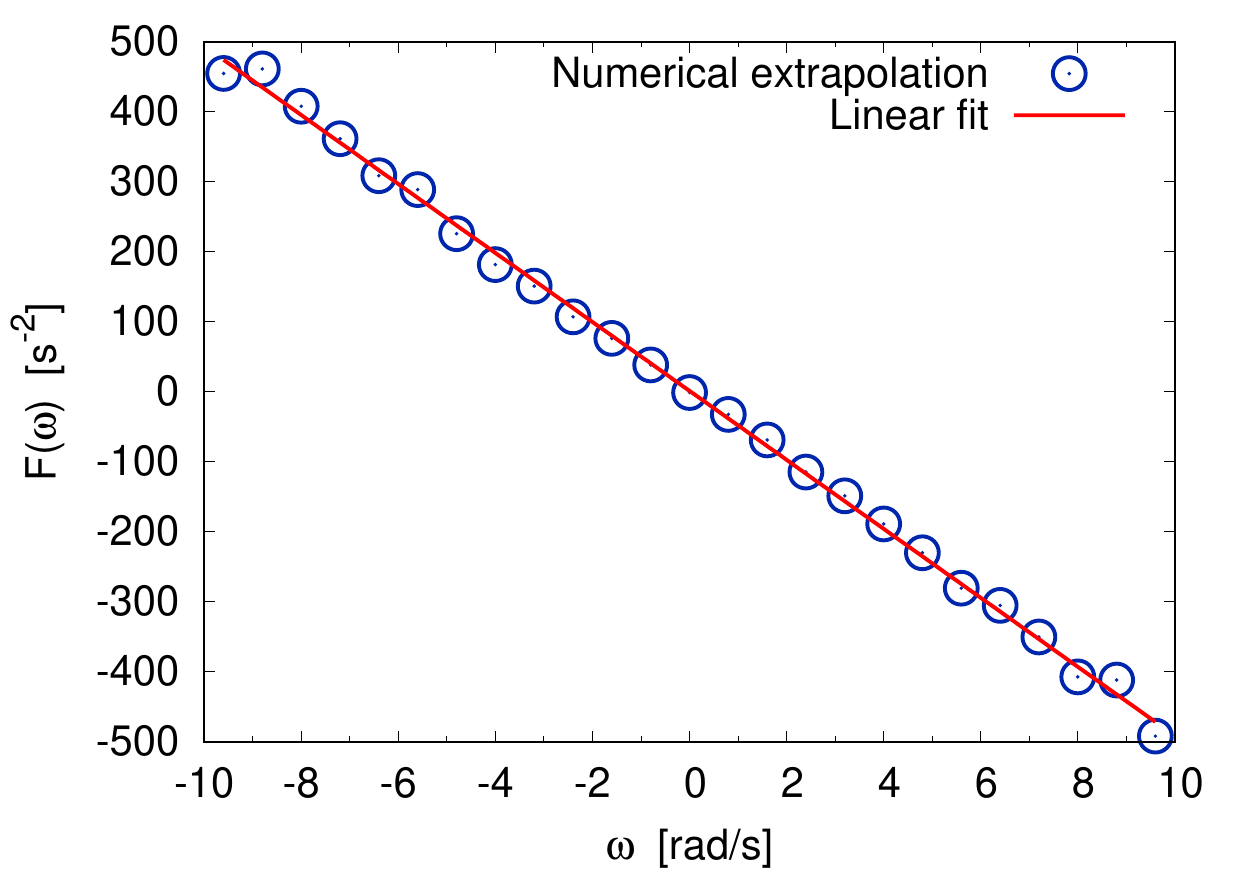}
 \includegraphics[width=0.49\linewidth]{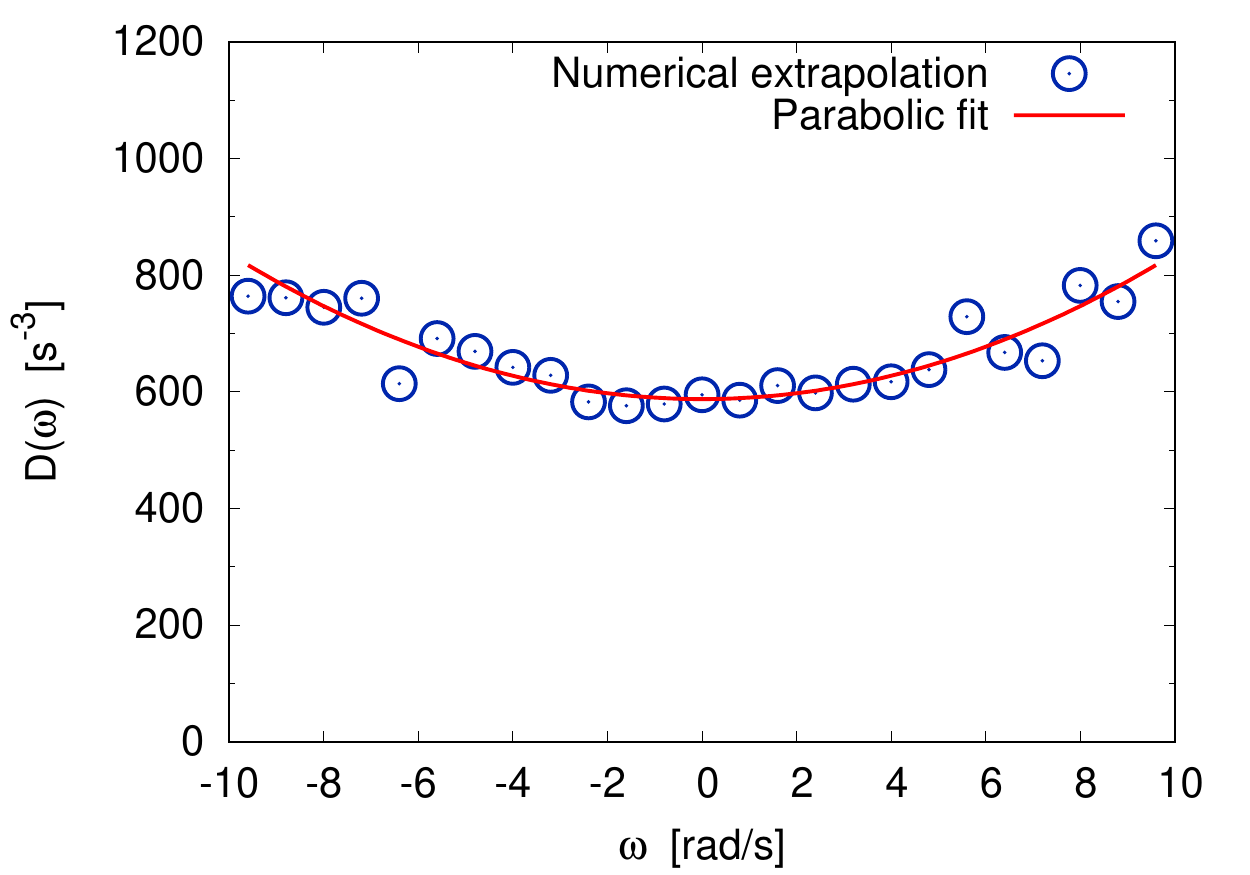}\\
 \includegraphics[width=0.49\linewidth]{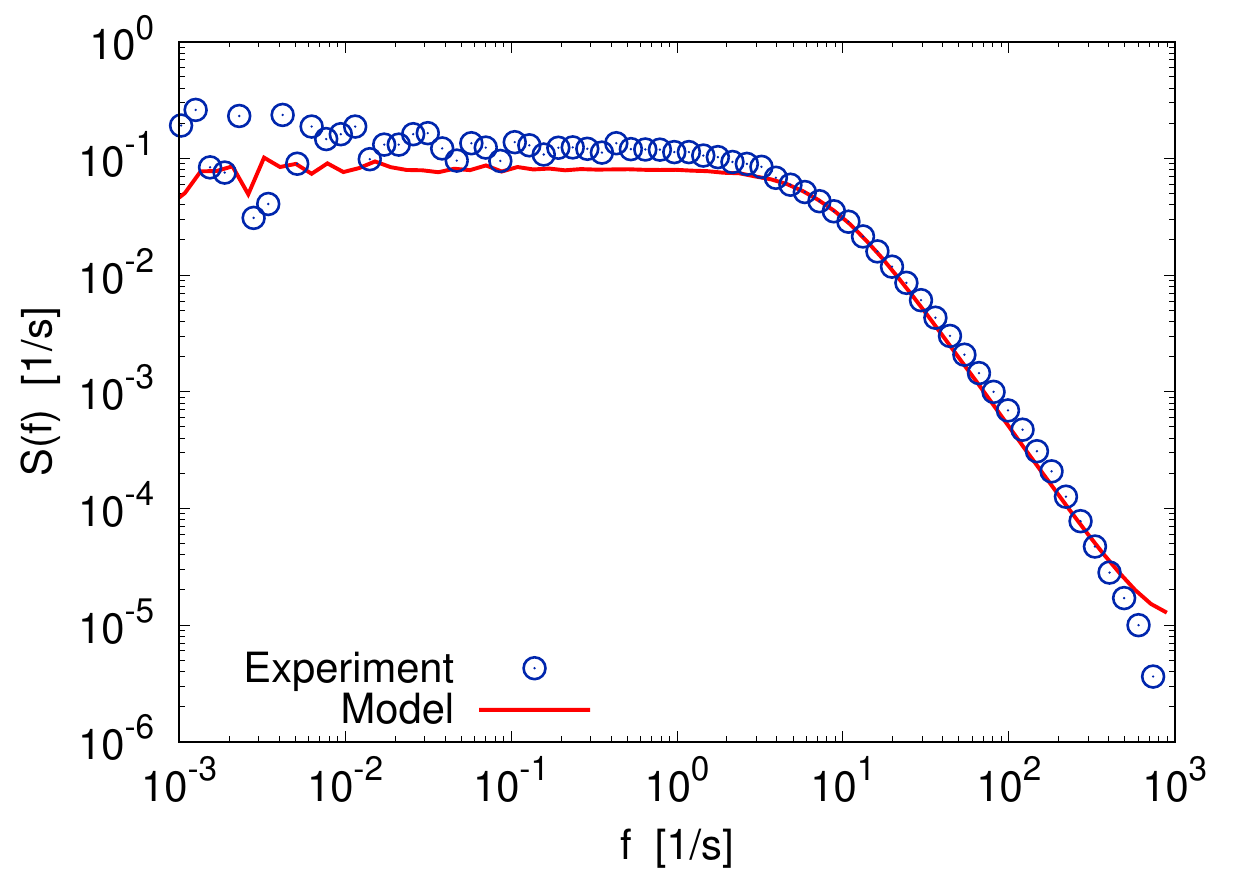}
 \includegraphics[width=0.49\linewidth]{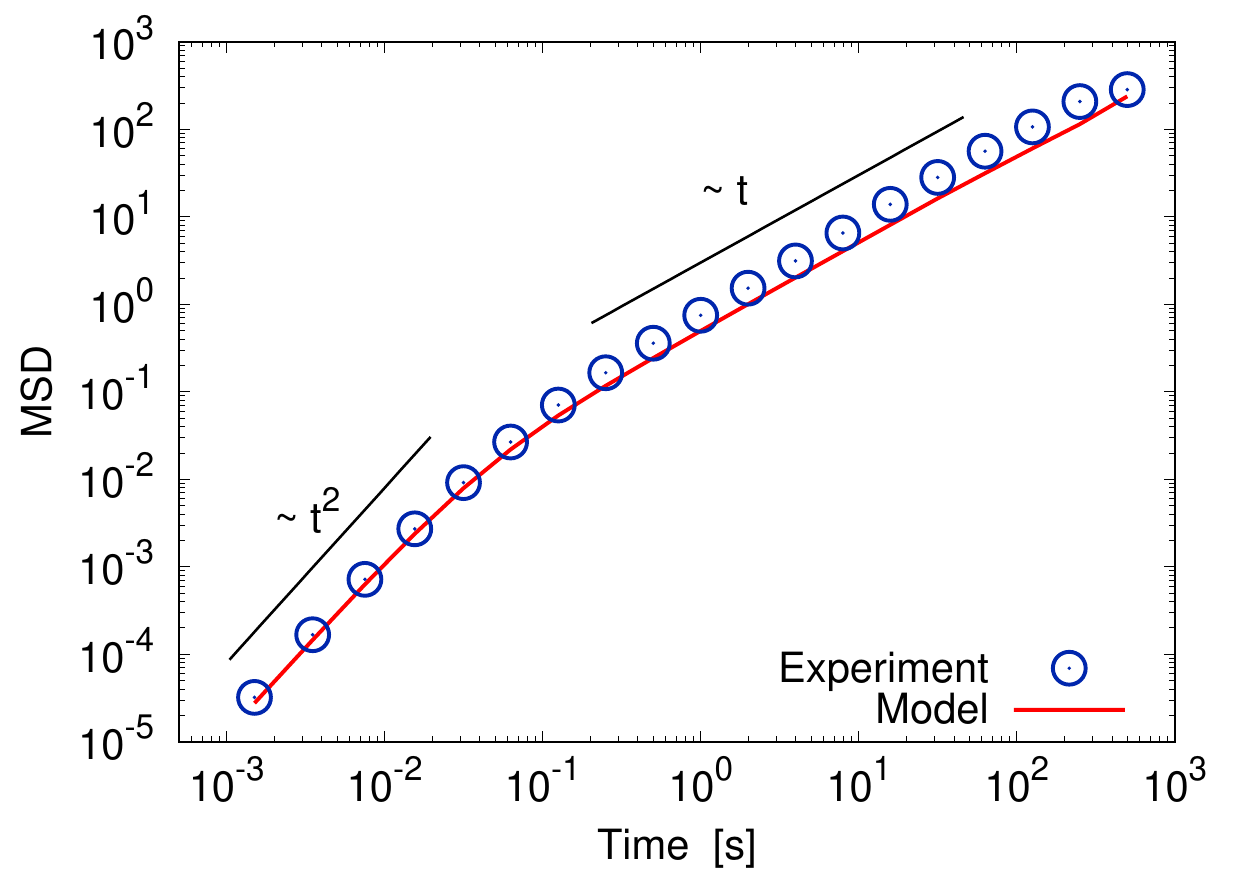}
 \caption{Dilute granular experiment.  Top: Reconstructed drift (\textbf{left})
   and diffusivity (\textbf{right}) for the angular velocity, fitted trough a
   linear function and a quadratic polynomial, respectively.  Bottom:
   Power spectrum (\textbf{left}) and mean square displacement (\textbf{right});
   experimental data (blue circles) are compared to those obtained by
   simulating the inferred stochastic model (red lines).  }
 \label{fig:experiment}
 \end{figure}

{\em Granular system} For the two previous examples we have a good
theoretical understanding, therefore the numerical procedure does not
reserve much surprises. More interesting is to use the procedure on
{\em experimental} data of a system which lacks a well established
theoretical understanding, for instance when far from equilibrium. The
analyzed time series is that of the angular velocity of a rotator
suspended in a vibrofluidized granular medium~\cite{scalliet}.  The
granular medium made of $N=300$ spheres of diameter $d=4$ mm is placed
in a cylindrical container of volume $\sim 7300$ times that of a
sphere (the average packing fraction is therefore $\sim$4\%). The
container is vertically shaken with a signal whose spectrum is
approximately flat in a range $[f_{min},f_{max}]$ with $f_{min}=200$
Hz and $f_{max}=400$ Hz.  A blade, our ``massive tracer'' with cross
section $\sim$$8d \times 4d$, is suspended into the granular medium and
rotates around a vertical axis. Its angular velocity $\omega(t)$ and
the traveled angle of rotation $\theta(t)=\int_0^t \omega(t')dt'$ are
measured with a time-resolution of $2$ kHz. The blade, interacting
with the spheres, performs a motion qualitatively similar to an
angular Brownian motion.  The shaking intensity is measured by the
normalized mean squared acceleration of the vibrating plate
$\Gamma=\sqrt{\langle (\ddot{z})^2 \rangle}/g \sim 40$. Again, in view
of the diluteness of the system and the large mass ratio between the
tracer and the particles, the dynamics of $\omega(t)$ is expected to
be well reproduced by a Markov process.  In Figure~\ref{fig:experiment}
we show that the damping force is proportional to the angular
velocity, as in the previous cases, while the reconstructed
diffusivity presents a parabolic dependence on $\omega$.  Such
multiplicative stochastic process can be numerically simulated and
compared to the experimental data.

The three examples above concern situations where, even if not totally
trivial, it is highly expected that a Markovian continuous model (a
LE) provides an effective description. More in general, however, the guess that $v$ is
described by a Markovian process is not trivial at all.  For instance,
the experiment discussed in the third case above has been conducted in
much more dense granular setups, for which there are several indications 
that the dynamics of the tracer is not
Markovian~\cite{scalliet,lasanta}. Basically we have that the variable
$\theta(t)$ is not able to describe in a good way the dynamics of the
tracer and it is necessary to introduce (at least) another
variable. Even the identification of such a variable is not obvious, this is part of a general problem which has been stressed by
Onsager and Machlup in their seminal work on fluctuations and
irreversible processes \cite{ons}, with the caveat: {\it How do you
  know you have taken enough variables, for it to be Markovian?}  In a
similar way, Ma noted that \cite{ma}: {\it The hidden worry of
  thermodynamics is: We do not know how many coordinates or forces are
  necessary to completely specify an equilibrium state.}

\section{Conclusions and Final Remarks \label{sec:conclusion}}
After a tribute to two classical models which had a seminal role in
science, namely the Lorenz model and the Lotka-Volterra equations, we
discussed the problem of model building and prediction with a
data-based approach.  A critical analysis shows the severe limits of
such an approach and the necessity to use, in a clever way, models.
Many interesting systems have a multiscale structure, i.e., there are
different variables with very different characteristic times.  While
such a feature is at the origin of severe difficulties, it also opens
the possibility to reduce the complexity by building an effective
equation for the slow variables. In this respect we discussed the
methodology and the difficulties to build effective equation for
proteins, as well as mesoscopic descriptions based on the Langevin
equation.  Reviewing some of the methodologies used in the building of
different models, it is well clear that there are no systematic
protocols and, on the contrary it is always necessary to use the
previous understanding of the considered system and, often, analogy
and intuition \cite{dah,hos}.

Another important point in model building, perhaps the most difficult
one, is to choose the ``right'' variables. This aspect, too often
overlooked, was touched while discussing the Langevin equation, and
underlined by Onsager and Machlup in their seminal paper \cite{ons}
and by Ma in his book on the foundation of statistical mechanics
\cite{ma}.  The ability of choosing the ``right''
  variables typically requires a conceptual abstraction which is key
  to scientific discoveries. For instance, in the '80s, some
  researchers in the field of artificial intelligence (AI) devised
  BACON, a computer program to automatize scientific discoveries~\cite{R1}
  (after the philosopher Francis Bacon who has been the
    champion of an inductive approach to science).  Apparently, BACON
  was able to ``discover'' the Black's law for temperature of a
  mixture of two liquids, the Snell's law in optics, and the third
  Kepler's law.  Looking at the details of the procedure used by
  BACON, however, it seems difficult to conclude that naive AI methods
  can replace the traditional creative approach to scientific
  discoveries.  Indeed, in the case of the third Kepler's law, BACON
  used as input the numerical values of distance from the Sun, $D$,
  and revolution period, $P$, of planets. The program, then,
  discovered that $D^3$ is proportional to $P^2$.  It is unfair to say
  that this represents a direct inductive approach only from data: The
  raw observables are not $D$ and $P$, but a list of planetary
  positions seen from the Earth at different times.  In his discovery,
  Kepler chose the ``right'' variables $D$ and $P$ as he was guided by
  strong beliefs in mathematical harmonies as well as the (at that
  time) controversial theory of Copernicus \cite{R2}.

We conclude with some general considerations about recent ideas on the
role of models and data in science.  In the last centuries the
scientific tradition had been based on two general pillars: A
theoretical one grounded on mathematics and experimental methods.
After World War II, numerical simulations surged as a third pillar for
scientific investigations. In the most interesting cases, simulations
are not a mere way to solve difficult mathematical problems but can
reach the status of experiments that can trigger new theoretical
developments. For example, the presence of slow decay in the molecular
velocity autocorrelation in liquids (hydrodynamic tails) has been
first observed in molecular dynamics simulations and then
theoretically explained \cite{hansen1990theory}.

In the last decade, the possibility of extracting knowledge by data
mining (i.e., through the algorithmic analysis of large amounts of
data) seems to suggest the emergence of a fourth paradigm, a new
scientific methodology to be added to the three already existing
\cite{hey}.  Ten years ago Chris Anderson, then the chief editor of
the influential technology magazine \emph{Wired},
published an article entitled ``The End of Theory: The Data Deluge
Makes the Scientific Method Obsolete'' \cite{and}.  This article
quickly became an ideological manifesto of datacentric enthusiasm
supporting a general philosophy starting from ``raw data'', without
constructing modeling hypotheses and, therefore, without
theory. We think that the surveyed procedures and, in
  particular, the difficulties of model building justify some skepticism
  about the claim that we are facing a new scientific revolution, the
datacentric one. However, it should be acknowledged that recent
results based on artificial intelligence (machine learning), as
discussed at the end of Section~\ref{sec:pred1}, applied to data seem
to support, at least in certain cases, the possibility either to
construct mathematical models or to improve their predictability
\cite{Lipson,brunton,Jaeger,Ott1}. Nevertheless, two remarks are in
order. The first one is that the above methods require the knowledge
of the state variables. Second, one has to restrict a priori the class
of mathematical structures (e.g., local interactions, degree of
nonlinearity, and symmetries, etc.) and this cannot be done
automatically without prior knowledge. Therefore, we can fairly
conclude that such approaches can be powerful tools at a technical
level but are not free from the key difficulties of model building.
\vspace{6pt}

\acknowledgments{We thank S. L. Brunton for interesting discussions
  about the problem of building equations from data.}
  
\appendix
\section{An attempt to build  clever low dimensional sysstems}

This Appendix briefly presents an interesting method that combines
mathematical models (i.e.,~the equations which are assumed to be known)
and experimental data to derive clever low dimensional models. This
can be seen as a way to improve standard Galerkin methods, such as
that discussed in Section~\ref{sec:lorenz} for deriving the Lorenz model.

In the presence of coherent structures, sometimes, the dimension of
the system is not too large so one can hope that few variables are
enough for a good description.  On the other, although from a
mathematical point of view the choice of the functions $\{ \phi_n \}$
is arbitrary, if one wants to use moderate values $N$ it is necessary
to use a ``clever'' complete, orthonormal set of eigenfunctions
\cite{Holmes2012}.

Of course one has to pay a price: The $\{ \phi_n \}$ must be selected
according to the specific dynamical properties of the problem under
investigation, and, in addition it is necessary to experimental or
numerical data. Let us discuss the method called proper orthonormal decomposition
(POD), which allows to derive clever low-dimensional systems.
  
The procedure is rather elaborated, in few words: One has to find the
eigenvalues and eigenfunctions of the integral equation
$$
\int R({\bf x}, {\bf x '}) \phi_n({\bf x'}) d {\bf x'}= \lambda_n \phi_n({\bf x})
$$ where the kernel $R({\bf x}, {\bf x '})$ is given by the spatial
correlation functions, which must be computed by experimental (or
numerical) data.  For instance, assuming, just for sake of notation
simplicity, that $\psi$ is a scalar field one has $R({\bf x}, {\bf
  x'})=\langle \psi({\bf x}) \psi({\bf x'})\rangle$.  The field
$\psi({\bf x},t)$ is thus reconstructed using this set of function and
the expansion (\ref{B2}).  The eigenvalues $\{ \lambda_n \}$ are
ordered in such a way to ensure that the convergence of the series is
optimal, i.e., this expansion is the best approximation in $L_2$ norm.

Essentially, the POD procedure is a special case of the Galerkin
method which captures the maximum amount of $L_2$ norm among all the
possible truncations with $N$ fixed.  Such an approach has been
successfully used to model different phenomena such as, e.g.,
jet-annular mixing layer, 2D flow in complex geometries and the
Ginzburg-Landau equation \cite{Holmes2012,smith2005low}.

We conclude by stressing that POD is not a straightforward procedure
as thoughtful selections of truncation and norms are
necessary in the construction of convincing low-dimensional models.

\bibliography{biblioA}

\end{document}